\documentclass[final]{IEEEtran}
\usepackage{cite}
\usepackage{graphicx}
\usepackage{picinpar}
\usepackage[cmex10]{amsmath}
\usepackage{amsmath,amsfonts,amssymb}
\usepackage{subfigure}
\usepackage{amsthm}
\usepackage{algorithmic}
\usepackage{stfloats}
\usepackage{bm}
\usepackage{color}
\usepackage{booktabs}
\usepackage{makecell}
\usepackage{mathrsfs}
\usepackage{stfloats}
\usepackage{url}
\usepackage{multirow}
\usepackage{threeparttable}
\usepackage{lipsum}
\usepackage[ruled,lined,boxed]{algorithm2e}
\usepackage{epstopdf}
\DeclareMathAlphabet{\mathbbold}{U}{bbold}{m}{n}

\begin{document}
\pdfoptionpdfminorversion=6
\newtheorem{lemma}{Lemma}
\newtheorem{corol}{Corollary}
\newtheorem{theorem}{Theorem}
\newtheorem{proposition}{Proposition}
\newtheorem{definition}{Definition}
\newcommand{\e}{\begin{equation}}
\newcommand{\ee}{\end{equation}}
\newcommand{\eqn}{\begin{eqnarray}}
\newcommand{\eeqn}{\end{eqnarray}}
\renewcommand{\algorithmicrequire}{ \textbf{Input:}} 
\renewcommand{\algorithmicensure}{ \textbf{Output:}} 

\title{Hybrid Knowledge-Data Driven Channel Semantic Acquisition and Beamforming for Cell-Free Massive MIMO}
\author{Zhen~Gao, Shicong~Liu,~\IEEEmembership{Graduate Student Member,~IEEE}, Yu~Su,~\IEEEmembership{Member,~IEEE}, Zhongxiang~Li, and Dezhi~Zheng~\IEEEmembership{Member,~IEEE}
	\thanks{
		
	This work was supported in part by the National Natural Science Foundation of China (NSFC) under Grant U2233216 and Grant 62071044, in part by the Shandong Province Natural Science Foundation under Grant ZR2022YQ62, and in part by the Beijing Nova Program (\textit{Corresponding author: Zhongxiang Li}).
	
	Shicong Liu is with the School of Information and Electronics, Beijing Institute of Technology, Beijing 100081, China.
	
	Zhen Gao, Zhongxiang Li, and Dezhi Zheng are with the MIIT Key Laboratory of Complex-Field Intelligent Sensing, Beijing Institute of Technology, Beijing 100081, China, also with Yangtze Delta
	Region Academy, Beijing Institute of Technology (Jiaxing), Jiaxing 314019, China, and also with the Advanced Technology Research Institute, Beijing Institute of Technology, Jinan 250307, China (e-mail: gaozhen16@bit.edu.cn).
	
	Yu Su is with China Mobile (Chengdu) Institute of Research and Development, Chengdu, Sichuan, 610000, China.
	}
}

\maketitle

\begin{abstract}
	This paper focuses on advancing outdoor wireless systems to better support ubiquitous extended reality (XR) applications, and close the gap with current indoor wireless transmission capabilities. We propose a hybrid knowledge-data driven method for channel semantic acquisition and multi-user beamforming in cell-free massive multiple-input multiple-output (MIMO) systems. Specifically, we firstly propose a data-driven multiple layer perceptron (MLP)-Mixer-based auto-encoder for channel semantic acquisition, where the pilot signals, CSI quantizer for channel semantic embedding, and CSI reconstruction for channel semantic extraction are jointly optimized in an end-to-end manner. Moreover, based on the acquired channel semantic, we further propose a knowledge-driven deep-unfolding multi-user beamformer, which is capable of achieving good spectral efficiency with robustness to imperfect CSI in outdoor XR scenarios. By unfolding conventional successive over-relaxation (SOR)-based linear beamforming scheme with deep learning, the proposed beamforming scheme is capable of adaptively learning the optimal parameters to accelerate convergence and improve the robustness to imperfect CSI. The proposed deep unfolding beamforming scheme can be used for access points (APs) with fully-digital array and APs with hybrid analog-digital array. Simulation results demonstrate the effectiveness of our proposed scheme in improving the accuracy of channel acquisition, as well as reducing complexity in both CSI acquisition and beamformer design. The proposed beamforming method achieves approximately $96\%$ of the converged spectrum efficiency performance after only three iterations in downlink transmission, demonstrating its efficacy and potential to improve outdoor XR applications.
\end{abstract}


\section{Introduction}
\label{S1}
\IEEEPARstart{T}{he} fifth generation (5G) mobile communication technology is envisioned to support diverse usage scenarios and applications, and has raised new challenges and opportunities, which promotes encouraging breakthroughs in recent years\cite{itu}. One of the most promising technologies is millimeter wave (mmWave) communication, which offers a significant increase in available spectrum resources, but also results in a substantial reduction in wireless coverage due to the high path loss of mmWave signals. Despite this challenge, the short wavelength of mmWave signals enables the deployment of massive multiple-input multiple-output (mMIMO)\cite{mMIMO} arrays with smaller physical space, which achieves superior performance in beamforming gain. {But the exponentially growing service demand such as the real-time video streaming}, and the increasing number of wireless user equipment (UE) nodes impose heavy burden on base stations (BSs) to provide balanced coverage. Due to severe inter-cell interference (ICI) in dense cellular networks\cite{cfcoverage}, spectrum efficiency (SE) is poor for cell edge UEs, which restricts further improvements.

To mitigate the ICI, signal co-processing concepts such as network MIMO\cite{netMIMO}, coordinated multi-point with joint transmission (CoMP-JT)\cite{CoMP}, and cloud radio access network (C-RAN)\cite{CRAN} were proposed. However, as long as the intrinsic cell-centric (or BS-centric) structure remains, interference among cells or clusters of coordinated cells are still non-negligible. Instead, as we consider transferring the cell-centric coordination to a user-centric fashion\cite{cfcoverage,CRAN}, i.e., the cell-free (CF) system, where each UE is served by its selected BSs, intra-cell interference can be precisely canceled among the selected cells. Moreover, as the cell clusters are dynamically selected by the UEs, cell boundaries naturally no longer exist, which simultaneously removes the ICI.

Acquiring downlink channel state information (CSI) at access points (APs) is of great significance to accomplish the ubiquitous coverage as well as high SE performance in XR applications. In conventional CF systems operating in time division duplex (TDD) mode, APs can obtain downlink CSI through uplink training thanks to the channel reciprocity that is commonly assumed in previous research\cite{cfcoverage}. Acquiring CSI instantaneously is a prerequisite for designing efficient beamforming. However, TDD requires strict synchronization to avoid interference and collision of uplink and downlink signals, and the same frequency band shared by uplink and downlink with time-division transmission would  lead to shorter coverage and lower UE mobility support than frequency division duplex (FDD) systems. FDD systems with dedicated uplink and downlink spectrum avoid such defects, while require excessive downlink training for accurate CSI acquisition as the perfect reciprocity no longer holds. In fact, uplink/downlink channel reciprocity may not always hold even in TDD systems due to radio frequency (RF) chain calibration error\cite{caliberr}, particularly for mmWave mMIMO systems with a large number of mmWave radio frequency chains.

\subsection{Related Work}

In recent years, downlink channel acquisition schemes based on sparse signal processing have received considerable attention. Common approaches can be broadly classified into CSI acquisition relying on  uplink/downlink channel reciprocity\cite{emilest,nanyang,ullearning,mxsjsac} and CSI estimation with feedback \cite{mxsjsac,FDD,dlcemumassiveMIMO_OMP,domp,dlcemumassiveMIMO,fddcf1}. The former category relies on the TDD assumption, e.g., \cite{emilest} considered perfect uplink/downlink CSI reciprocity, and \cite{nanyang} proposed a compressive sensing (CS)-based method to fully utilize the delay domain sparsity as auxiliary information to reduce the channel estimation overhead. Additionally, learning-based uplink CSI acquisition has recently garnered significant attention due to its demonstrated superior performance and reduced overhead\cite{ullearning,mxsjsac}. 

However, the reciprocity in TDD systems does not always hold, particularly for mmWave mMIMO systems\cite{caliberr}, which motivates researchers to focus more on the research of downlink CSI acquisition and feedback\cite{FDD,dlcemumassiveMIMO_OMP,domp,dlcemumassiveMIMO,fddcf1}. Specifically, \cite{FDD} utilized the common sparse support of angle-domain mMIMO channels among different subcarriers, 
and requires manually adjusted thresholds to achieve better performance. Joint orthogonal match pursuit (JOMP) proposed in \cite{domp} has assumed a large proportion of common sparse support of angle-domain mMIMO channels shared by multiple adjacent UEs, but the authors have not considered the UE power imbalance due to large-scale fading in practical scenarios. The aforementioned papers have also neglected the angular reciprocity in FDD systems, which can effectively improve the reconstruction accuracy. A two-step iteration method is proposed in \cite{dlcemumassiveMIMO_OMP} for downlink channel reconstruction, which has taken the angular reciprocity in FDD systems into consideration. However, the gradient-based methods heavily depend on their initial values and present prohibitively high computational complexity at the stage of online estimation. A sparse reconstruction algorithm relying on the vector approximate message passing (VAMP) algorithm is proposed in \cite{ampest1} under a specific multivariate Bernoulli-Gaussian {\em a priori} assumption. Moreover, generalized AMP (GAMP)-based method is proposed to adapt to non-linear processing such as quantization, which also no longer requires accurate {\em a priori} distribution\cite{cellfreekml}. Nevertheless, AMP-based methods still require a large number of iterations to achieve convergence, leading to high computational complexity.

In mMIMO systems with OFDM modulation, the feedback overhead that scales linearly with the numbers of antennas and subcarriers is another burden to concern. Fortunately, deep learning (DL)-based CSI feedback approach emerges as one of the most promising solutions to further reduce the feedback overhead and complexity\cite{csinet,selfinfofb,MCfb,feedbackdl}. Since the fully connected (FC) layers have excessively large amount of parameters in image-like data processing tasks, convolutional neural networks (CNN)-based methods are proposed to reduce the number of parameters and fully utilize the spatial correlation properties~\cite{csinet,feedbackdl}. However, CNN-based models have limited receptive fields, and the amount of computation scales linearly with the number of image pixels\cite{attn0} as well as the amount of layers. These drawbacks would lead to restricted performance in CSI acquisition and impose computational burdens on UEs when processing large-scale CSI data with sophisticated features. To further improve the feedback accuracy, we may consider incorporating additional features manually into the design of the neural network architecture\cite{selfinfofb} or exploring other mechanisms for automatically extracting more features\cite{attn,mlpmixer}. \cite{transfb} proposed to feedback downlink CSI via a full attention network, which outperforms CNN-based neural network (NN) architectures, but the computational complexity in floating point operations per second (FLOPS) and the number of parameters is extremely high due to the involved multi-head attention module.

In recent years, the number of antennas in wireless communication has been increasing, which has posed significant challenges to beamforming algorithms. Among them, wireless communication systems based on ultra large-scale arrays will suffer from severe beam squint effect\cite{TTD}, especially in ultra wideband systems, which significantly increases the beam training overhead and causes power leakage and interference. To eliminate such effect, researchers in \cite{TTD} proposed to fully utilize true time delay (TTD) to achieve frequency-dependent phase shifts. However, it is less likely to introduce beam squint effect in CF-mMIMO systems with distributed antenna configuration~\cite{distiter,LPZF,fddcf1}. Authors in~\cite{LPZF} have proposed multiple distributed zero-forcing (ZF)-based beamforming schemes under several ideal assumptions such as infinite front-haul capacity and Rayleigh channel. \cite{distiter} has  considered a relatively practical propagation environment, and has introduced over-the-air signaling mechanism to cut off the backhaul signaling requirements. However, the aforementioned papers have only considered TDD-based CF system with uplink estimation. Under the fact that imperfect reciprocity exists in both TDD and FDD mMIMO systems, downlink channel estimation and CSI feedback would be preferred as a unified paradigm. As far as we concern, only a few papers such as \cite{fddcf1} have proposed to obtain the downlink CSI by feeding dominant path gain back to the BS.

\subsection{Our Contributions}
In this paper, we propose a learning-based channel semantic acquisition and beamforming solution for CF-mMIMO systems. The proposed approach can significantly reduce the CSI acquisition overhead, lower the computational amount of UE, and improve robustness to imperfect CSI. We firstly introduce a novel CSI estimation and feedback auto-encoder network adopting new network architecture, 
and train the network with goal-oriented cosine-similarity loss. The proposed network structure and training policy substantially reduce the computational latency, which is much lower than the conventional CNN, and obtains lower reconstruction error. Moreover, with more accurately reconstructed downlink CSI at APs, we propose to unfold conventional linear beamforming scheme with deep learning, which shows lower complexity than the matrix inversion-based methods in both fully-digital array and hybrid analog-digital array. In summary, our contributions are as follows
\begin{itemize}
	\item \textbf{Novel network architecture for pilot design, downlink estimation, and CSI feedback in CF-mMIMO systems}. Conventional CNN-based feature-extraction models suffer from high computational latency which is linearly proportional to the dimension of CSI to be feedback, and occupies huge memory for storage and computational resources. Therefore, we propose a data-driven channel estimation (or embedding) scheme with multi-layer perceptron (MLP)-Mixer\cite{mlpmixer} module. The proposed scheme requires no successively sliding convolution operations, and introduces patching-embedding mechanism as well as mixer-layer computation, which can draw global dependencies rather than the limited receptive fields in CNN-based models.
	\item \textbf{Goal-oriented reconstruction target function in channel semantic extraction for beamforming design}. As perfect uplink/downlink channel reciprocity does not hold due to the different uplink/downlink frequency bands in FDD mode and non-negligible RF chain calibration error in high-frequency TDD mode, downlink estimated CSI needs to be fed back to APs for beamforming design. We proposed to use the cosine-similarity function as the the goal-oriented loss for beamforming design, which can significantly reduce the feedback overhead and improve the spectrum efficiency in beamforming.
	\item \textbf{Deep unfolding based robust beamforming schemes for both fully-digital array and hybrid analog-digital array}.
	Substantially reduced estimation and feedback overhead would inevitably lead to CSI reconstruction error at APs. Traditional channel inversion beamforming methods are sensitive to numerical disturbance, which severely degrades downlink SE performance. We hence propose a deep unfolding-based beamforming scheme that unfolds the successive over-relaxation (SOR)-based linear beamforming scheme, which can not only reduce the complexity but also improve the robustness by integrating learnable hyper parameters.
\end{itemize}

{\em Notations}: We use normal-face letters to denote scalars, lowercase (uppercase) boldface letters to denote column vectors (matrices). The $k$-th row vector, and the $m$-th column vector of matrix ${\bf H}\in\mathbb{C}^{K\times M}$ are denoted as ${\bf H}_{k,:}$ and ${\bf H}_{:,m}$, respectively, ${\bf H}_{\{k,m\}}$ denotes the element in the $k$-th row and the $m$-th column of ${\bf H}$, and $\{{\bf H}_n\}_{n=1}^N$ denotes a matrix set with the cardinality of $N$. ${\bf I}_{M\times N}$, ${\bf 1}_{M\times N}$ and ${\boldsymbol{0}}_{M\times N}$ denote the identity matrix, all-one matrix and zero matrix of size $M\times N$, respectively. The superscripts $(\cdot)^{\rm T}$, $(\cdot)^*$, and $(\cdot)^{\rm H}$ represent the transpose, conjugate, and conjugate transpose operators, respectively. We use $\otimes$ and $\odot$ to denote Kronecker product and Hadamard product, respectively. $\mathcal{CN}(\mu,\sigma)$ denotes the complex Gaussian distribution with mean $\mu$ and standard deviation $\sigma$. $\mathbb{E}[\cdot]$ denotes the statistical expectation operator.

\begin{figure*}[t]
	\centering
	\includegraphics[width=0.85\textwidth]{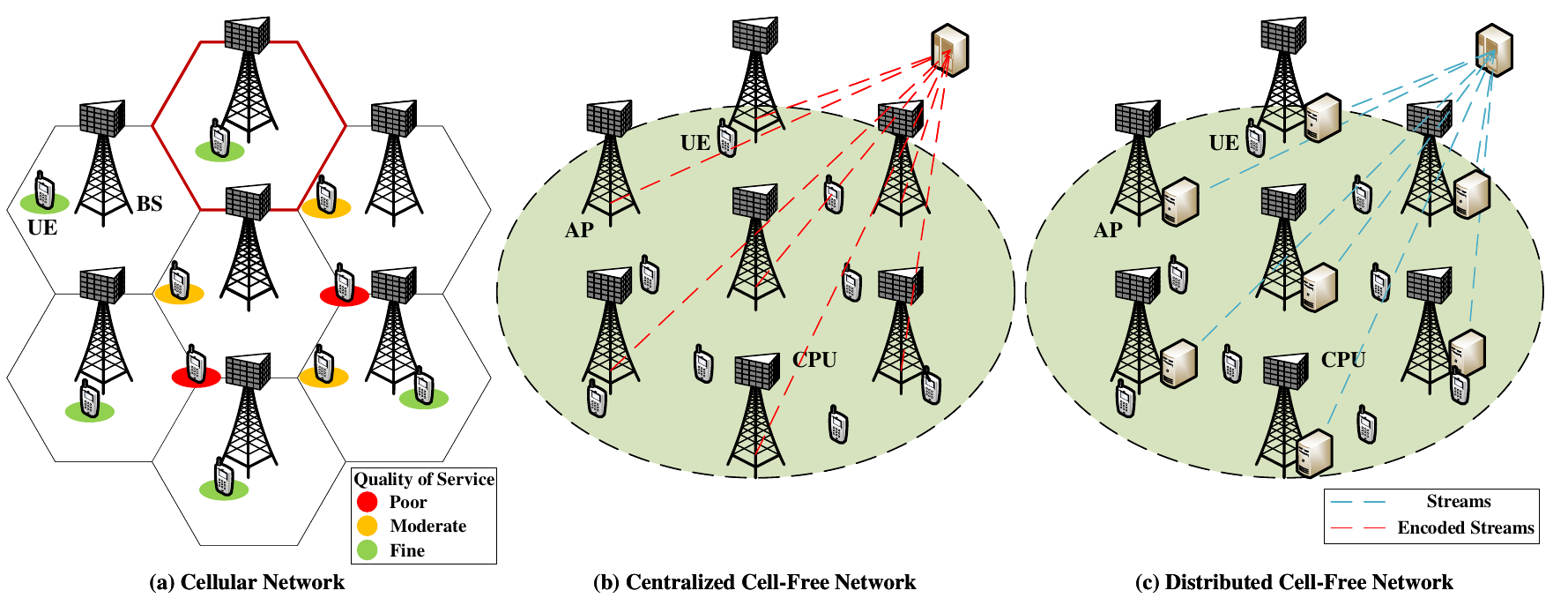}
	\caption{Schematic diagram of (a) conventional cellular network (b) centralized CF network and (c) distributed CF network. Users at cell edges in (a) suffers from poor quality of service (QoS) due to pilot contamination and inter-cell interference, which no longer exists (or can be significantly inhibited) in CF systems (b) and (c). {For scenario (b) the transmitted signals are pre-processed at CPU node, while for scenario (c) the transmitted signals are processed locally at APs after scheduling.}}\label{sysmodel}
\end{figure*}

\section{System Model}
Consider a typical CF-mMIMO system with $N$ APs, where each AP is equipped with a hybrid analog-digital array with $N_{\rm AP}$ antennas and $N_{\rm RF}$ RF chains. $U$ UEs, each with an $N_{\rm UE}$-element array to support single data stream, are cooperatively served by $N$ APs as illustrated in Fig. \ref{sysmodel}. It is assumed that $UN_{\rm UE}\leq N N_{\rm AP}$.

\subsection{Uplink Channel Estimation Procedure}\label{sec:uplink}
In the uplink transmission procedure, all $U$ users simultaneously transmit their assigned pilot signals {$ {\bf S}_{u} \in\mathbb{C}^{N_{\rm UE}\times \tau_p^{\rm UL}}, u\in\mathcal{U}=\{ u\mid 1\leq u \leq U, u\in\mathbb{Z} \}$} with a length of $\tau_p^{\rm UL}$. In conventional uplink channel estimation schemes, different UEs' pilots are mutually orthogonal, i.e.,
\begin{align}
	{\bf S}_{i}{\bf S}^{\rm H}_{j}={\bf 0}_{N_{\rm UE}\times N_{\rm UE}},\ \  \forall i,j\in\mathcal{U}, \ i\neq j,\notag
\end{align}
where ${\bf S}_{i}$ denotes the pilot sequence from the $i$-th UE, and the pilot signal is normalized by transmit power $p_u$, i.e., $\Vert {\bf S}_{u} \Vert^2_2=p_u\tau_p^{\rm UL} $. Therefore, the received signal at the $n$-th AP on the $k$-th subcarrier can be given as
\begin{equation}
	{\bf Y}^{\rm UL}_n[k] = \left( {\bf F}_n^{\rm RF}{\bf F}_n^{\rm BB}[k] \right)^{\rm H}\left(\sum_{u=1}^{U}{\bf H}_{u,n}^{\rm UL}[k]{\bf S}_{u} +{\bf N}_n[k]\right),
\end{equation}
where ${\bf Y}^{\rm UL}_n[k] \in\mathbb{C}^{N_{\rm S}\times \tau_p^{\rm UL}}$, ${\bf F}_n^{\rm RF}\in\mathbb{C}^{N_{\rm AP}\times N_{\rm RF}}$ and ${\bf F}_n^{\rm BB}[k]\in\mathbb{C}^{N_{\rm RF}\times N_{\rm S}}$ are respectively the analog part and digital part of the hybrid combiner at the $n$-th AP, $N_{\rm S}$ denotes the number of data streams at each AP, ${\bf H}_{u,n}^{\rm UL}{[k]}\in\mathbb{C}^{N_{\rm AP}\times N_{\rm UE}}$ is the channel matrix from the $u$-th UE to the $n$-th AP {on the $k$-th subcarrier}, and ${\bf N}_n{[k]}\in\mathbb{C}^{N_{\rm AP}\times \tau_p^{\rm UL}}$ denotes the additive white Gaussian noise (AWGN) matrix {of the $n$-th AP on the $k$-th subcarrier}. On this basis, we can formulate the uplink channel estimation for the $u$-th UE as 
\begin{equation}
	{\bf Y}_{u,n}^{\rm UL}[k] = {\bf Y}_{n}^{{\rm UL}}[k]{\bf S}_{u}^{\rm H}=\left(\tilde{\bf F}_n[k] \right)^{\rm H}{\bf H}_{u,n}^{\rm UL}[k]+\tilde{\bf N}_n[k],
		\label{eq:ulprocedure}
\end{equation}
where ${\bf Y}^{\rm UL}_{u,n}[k] \in\mathbb{C}^{N_{\rm S}\times N_{\rm UE}}$, $\tilde{\bf F}_n[k] = {\bf F}_n^{\rm RF}{\bf F}_n^{\rm BB}[k]\in\mathbb{C}^{N_{\rm AP}\times N_{\rm S}}$ denotes the effective hybrid combiner, and $\tilde{\bf N}_n[k] = {\bf N}_n[k]{\bf S}_{u}^{\rm H}$ is the effective AWGN matrix. By stacking a sufficient number of measurements ${\bf Y}_{u,n}^{\rm UL}[k]$ from multiple time slots\footnote{Eq.~\eqref{eq:ulprocedure} only shows the measurement model in one time slot and ignores the subscript of time slots for brevity. Details for multiple time slot concatenation can be found in~\cite{mxsjsac}, Eq. (3)-(4).}, we can estimate the uplink CSI ${\bf H}^{\rm UL}_{u,n} \in\mathbb{C}^{N_{\rm AP}\times N_{\rm UE}}$ for $u=1,\cdots,U$.

\subsection{Downlink Transmission}\label{sec:downlink}
In the downlink transmission of FDD systems, downlink channel estimation is necessary as the channel reciprocity assumed in TDD systems does not hold. For the mostly adopted assumptions in CF-mMIMO systems that transmit signals are jointly designed at central processing unit (CPU) as shown in Fig.~\ref{sysmodel}(b), the desired received downlink signal at the $u$-th UE on the $k$-th subcarrier can be expressed as
\begin{equation}
	y_u[k] = {\bf w}_u^{\rm H}[k]\left( \sum_{n=1}^{N} {\bf H}_{u,n}^{\rm DL}[k]{\bf D}_{u,n}{\bf F}_{n}^{\rm cen}[k] {\bf s}[k]+{\bf n}_u[k] \right),
\end{equation}
where {${\bf s}[k]\in\mathbb{C}^{U}$ is the transmitted signal}, ${\bf w}_u [k]\in\mathbb{C}^{N_{\rm UE}}$ is the $u$-th UE's combiner, ${\bf H}^{\rm DL}_{u,n}[k]\in\mathbb{C}^{N_{\rm UE}\times N_{\rm AP}}$ denotes the channel matrix from the $n$-th AP to the $u$-th UE, ${\bf F}^{\rm cen}_{n}[k] ={\bf F}^{\rm RF}_n{\bf F}^{\rm BB}_{n}[k]\in\mathbb{C}^{N_{\rm AP}\times U}$ is the part of the complete hybrid beamformer at the CPU associated with the $n$-th AP, and ${\bf n}_u[k]\in\mathbb{C}^{N_{\rm UE}\times 1}$ is the AWGN vector with ${\bf n}_u[k]\sim {\mathcal{CN}}({\bf 0},\sigma_u^2{\bf I}_{N_{\rm UE}\times N_{\rm UE}})$. We denote the user scheduling matrix as ${\bf D}_{u,n}\in\mathbb{Z}^{N_{\rm AP}\times N_{\rm AP}}$, which gives the association between $N$ APs and $U$ UEs. 
For example, when the $n$-th AP is assigned to serve the $u$-th UE, we have ${\bf D}_{u,n} = {\bf I}_{N_{\rm AP}}$, otherwise ${\bf D}_{u,n} = {\bf 0}_{N_{\rm AP}}$. The desired signal can be written in a compact form as
\begin{equation}
	y_u[k] = {\bf w}_u^{\rm H}[k]\left( {\bf H}_{u}^{\rm DL}[k] {\bf D}_u{\bf F}^{\rm cen}[k]{\bf s}[k]+{\bf n}_u[k] \right),
\end{equation}
where ${\bf H}_{u}^{\rm DL}=[{\bf H}_{u,1}^{\rm DL},\cdots,{\bf H}_{u,N}^{\rm DL}]\in\mathbb{C}^{N_{\rm UE}\times NN_{\rm AP}}$ is the channel matrix of the $u$-th UE, ${\bf D}_u={\rm blkdiag}({\bf D}_{u,1},\cdots,{\bf D}_{u,N})$ is the block diagonal scheduling matrix, and ${\bf F}^{\rm cen}=[({\bf F}_{1}^{\rm cen})^{\rm H},\cdots,({\bf F}_{N}^{\rm cen})^{\rm H}]^{\rm H}\in\mathbb{C}^{NN_{\rm AP}\times U}$ is the beamformer designed at CPU. 

In the centralized processing scheme, APs work as remote radio heads that transmit centrally processed signals from CPU. While this approach has the optimal beamforming performance, it also incurs excessive front-hauling overhead and prohibitive signal processing complexity, where both of them are proportional to the number of APs. Therefore, a distributed deployment and processing paradigm is preferred, where signals are firstly scheduled to different APs and then used for local beamforming design at each AP. The received downlink signal from $N$ distributed APs to the $u$-th UE on $k$-th subcarrier can be given as
\begin{equation}
	\begin{aligned}
		y_u[k] &= {\bf w}_u^{\rm H}[k]\left(\sum_{n=1}^{N} {\bf H}_{u,n}^{\rm DL}[k] {\bf F}_n^{\rm dis}[k]\tilde{\bf D}_n{\bf s}[k] + {\bf n}_u[k] \right)\\
		&={\bf w}_u^{\rm H}[k] \left( {\bf H}_{u}^{\rm DL}[k]\tilde{\bf F}^{\rm dis}{\bf s}[k] + {\bf n}_u[k] \right),
		\label{eq:dis}
	\end{aligned}
\end{equation}
where $\tilde{\bf D}_n=[{\bf d}_{{\mathcal U}_n^{1}},\cdots, {\bf d}_{{\mathcal U}_n^{\vert {\mathcal U}_n\vert}}]^{\rm T}\in\mathbb{Z}^{\vert {\mathcal U}_n\vert\times U}$ is the scheduling matrix, $\mathcal{U}_n$ is the UE index set associated with the $n$-th AP, and ${\bf d}_{i}$ with $ i\in {\mathcal{U}}_n$ is a binary vector with the $i$-th single element set to $1$ and all other elements set to $0$. $\vert {\mathcal{U}}_n \vert\leq U$ is the number of UEs served by the $n$-th AP, ${\bf F}_n^{\rm dis}\in\mathbb{C}^{N_{\rm AP}\times \vert {\mathcal{U}}_n \vert}$ is the local beamforming matrix at the $n$-th AP, and 
\begin{equation}
	\tilde{\bf F}^{\rm dis} = \left[\begin{array}{c}
		{\bf F}_1^{\rm dis}\tilde{\bf D}_1  \\
		{\bf F}_2^{\rm dis}\tilde{\bf D}_2  \\
		\vdots  \\
		{\bf F}_N^{\rm dis}\tilde{\bf D}_N
	\end{array}\right]\in\mathbb{C}^{NN_{\rm AP}\times U}
\end{equation}
is the equivalent distributed beamforming matrix. 
Without loss of generality, in this paper, we consider that all APs cooperatively serve all UEs in the centralized processing paradigm, which allows us to ignore the scheduling matrix ${\bf D}_u$ for convenience. We also assume that each AP is assigned $\vert {\mathcal{U}}_n \vert = N_{\rm S}$ UEs, $\forall\ 1\leq n\leq N$, in distributed processing paradigm\footnote{In this work, we do not consider the design of the scheduling matrix, and therefore it is sufficient for $\tilde{\bf D}_n$ to satisfy ${\rm rank}(\tilde{\bf D})=U$, where $\tilde{\bf D}=[\tilde{\bf D}_1^T,\cdots,\tilde{\bf D}_{N}^T]^T\in\mathbb{Z}^{NN_{\rm S}\times U}$.}.
\subsection{Channel Model}
Taking the downlink transmission as an example, the channel matrix from the $n$-th AP to the $u$-th UE in the delay domain can be expressed as
\begin{equation}
	\begin{aligned}
		{\bf G}_{u,n}^{\rm DL}(t)=&\frac{1}{\sqrt{L_pL_c}}\!\!\sum_{\ell=1}^{L_c}\sum_{p=1}^{L_p}\!\alpha_{\ell,p}^{\rm DL} \beta_{\ell,p}p(tT_s-\tau_{\ell,p})\\&{{\bf a}_{\rm R}(\phi_{n,\ell,p},\psi_{n,\ell,p}) {\bf a}_{\rm T}^{\rm H}(\phi^{\prime}_{u,\ell,p},\psi^{\prime}_{u,\ell,p})},
		\label{eq:chmod}
	\end{aligned}
\end{equation}
where $L_c$ and $L_p$ denote the number of scatter clusters and the number of multi-path components (MPCs) in each cluster, respectively\footnote{Without loss of generality, we assume that all UEs and APs have the same parameter configuration $L_c$ and $L_p$, which can be adjusted in simulations or practice as needed.}. {$\phi_{n,\ell,p}$ ($\phi^{\prime}_{u,\ell,p}$) and $\psi_{n,\ell,p}$ ($\psi^{\prime}_{u,\ell,p}$)} are respectively the azimuth and elevation angle at AP (UEs). $p(\cdot)$ is the pulse shaping function, where $T_s$ denotes the sampling period. $\tau_{\ell,p}$ is the discrete delay of the $p$-th path from the $\ell$-th cluster, and $\alpha_{\ell,p}^{\rm DL}\sim \mathcal{CN}(0,1)$ and $\beta_{\ell,p}=10^{-{\rm PL}_{\ell,p}/10}$ are the corresponding Rayleigh fading factor and large-scale fading factor\cite{channel3GPP}, respectively. Therefore, the frequency domain channel on the $k$-th subcarrier can be expressed as
\begin{align}
	{\bf H}^{\rm DL}_{u,n}[k]=\sum_{q=0}^{K-1}{\bf G}^{\rm DL}_{u, n}[q] e^{-j\frac{2\pi k}{K}q},
\end{align}
where ${\bf G}^{\rm DL}_{u,n}[q]={\bf G}^{\rm DL}_{u,n}(qT_s)$ is the discrete delay domain channel impulse response, and $K$ is the number of subcarriers. 
\section{Data-Driven CSI Acquisition based on Channel Semantic Embeddings}
\label{CEFB}
As we have mentioned in Section II, the uplink CSI obtained at each AP is only limited to their respective local CSI, while the downlink channel estimated at each UE can include the complete CSI associated with all $N$ APs, which can be used for centralized processing. In this section, we firstly formulate the CSI acquisition problem and propose a data-driven CSI acquisition approach, which can reconstruct the downlink channels from downlink pilot signals (with compressed observations) and the uplink feedback (with quantized channel semantic embedding). Specifically, for uplink transmission, the proposed learning-based estimation method can significantly reduce the uplink pilot overhead (i.e., observation dimension) and reconstruction error. While for the downlink, the proposed methods effectively extract the channel semantics, and can be optimized by a goal-oriented loss function. Additionally, we also discuss the characteristics of the adopted network backbone.

\subsection{Problem Formulation}

\label{sec:formulation}
Consider a CF-mMIMO system operating in FDD mode. At the stage of downlink pilot transmission, the required number of time slots is $\tau_p^{\rm DL}\tau_{\rm UE}$, which includes $\tau_p^{\rm DL}$ transmit time blocks and each time block includes $\tau_{\rm UE}$ receive time slots. Specifically, in the $\tau$-th time slot of the $\tau^\prime$-th time block, the received pilot signal at the $u$-th UE can be given as
\begin{equation}
	{y}_{u,\tau,\tau^\prime}^{\rm DL}[k] = {\left( {\bf w}_{u,{\tau}} [k]\right)}^{\rm H}\left( {\bf H}_{u}^{\rm DL}[k] \tilde{\bf f}_{\tau^\prime}[k] + {\bf n}_{u,\tau,\tau^\prime}[k] \right),
	\label{eq:pilottrain1}
\end{equation}
where ${ {\bf w}_{u,\tau} }[k] \in\mathbb{C}^{N_{\rm UE}}$ is the $\tau$-th ($\tau = 1,\cdots,\tau_{\rm UE}$) receive time slot's combining vector at the $u$-th UE, and $\tilde{\bf f}_{\tau^\prime}[k] = \tilde{\bf F}^{\rm cen}_{\tau^\prime}[k]{\bf s}_{\tau^\prime}[k]\in\mathbb{C}^{NN_{\rm AP}\times 1}$ or $\tilde{\bf F}^{\rm dis}_{\tau^\prime}[k]{\bf s}_{\tau^\prime}[k]\in\mathbb{C}^{NN_{\rm AP}\times 1}$ is the pilot signal vector in the $\tau^\prime$-th ($\tau^\prime=1,\cdots,\tau_p^{\rm DL}$) transmit time block. Assuming frequency-flat pilots for brevity\footnote{To avoid potential high peak-to-average power ratio cause by this assumption, we can introduce a predefined frequency-domain pseudo-random scrambling code {across different subcarriers according to~\cite{scrambling}}.}, by collecting the received pilot signals in $\tau_{\rm UE}\tau_p^{\rm DL}$ time slots, we can obtain the measurement matrix as
\begin{equation}
	\tilde{\bf Y}_u^{\rm DL}[k] =  {\bf W}_u^{\rm H} \left( {\bf H}_{u}^{\rm DL}[k]\tilde{\bf F}+ {\bf N}_u[k] \right) ,
	\label{eq:pilottrain2}
\end{equation}
where the element in the $\tau$-th row and the $\tau^\prime$-th column of $\tilde{\bf Y}_u^{\rm DL}[k]\in\mathbb{C}^{\tau_{\rm UE}\times \tau_p^{\rm DL}}$ is ${y}_{u,\tau,\tau^\prime}^{\rm DL}[k]$, ${\bf W}_u^{\rm H}=[{\bf w}_{u,1},\cdots,{\bf w}_{u,\tau_{\rm UE}}]^{\rm H}\in\mathbb{C}^{\tau_{\rm UE}\times N_{\rm UE}}$ is the stacked combining matrix, and $\tilde{\bf F}=[\tilde{\bf f}_1,\cdots,\tilde{\bf f}_{\tau_p^{\rm DL}}]\in\mathbb{C}^{NN_{\rm AP}\times \tau_p^{\rm DL}}$ is the aggregated pilot signal matrix. We can easily rewrite the pilot training process for the $u$-th UE in $\tau_{\rm UE}\tau_p^{\rm DL}$ slots as
\begin{equation}
	\tilde{\bf y}_u^{\rm DL}[k] = {\rm vec}\left(\tilde{\bf Y}_u^{\rm DL}[k]\right) =  {\bf P}\: {\rm vec}\left( {\bf H}_{u}^{\rm DL}[k] \right) + \tilde{\bf n}_u[k],
	\label{eq:pilottrain3}
\end{equation}
where ${\bf P} = \tilde{\bf F}^{\rm T}\otimes {\bf W}_u^{\rm H}\in\mathbb{C}^{\tau_{\rm UE}\tau_p^{\rm DL}\times NN_{\rm AP}N_{\rm UE} }$ is the overall pilot matrix, and $\tilde{\bf n}_u[k]={\rm vec}({\bf W}_u^{\rm H}{\bf N}_u[k])\in\mathbb{C}^{\tau_{\rm UE}\tau_p^{\rm DL}}$ is the vertorized overall AWGN matrix. Recall the channel model in Eq.~\eqref{eq:chmod}, channel matrix can be further given in a compact form as ${\bf H}_{u,n}^{\rm DL}[k] = {\bf A}_{\rm R} {\bf H}_{u,n}^{\rm G,DL}[k]{\bf A}_{\rm T}^{\rm H}$, where ${\bf A}_{\rm R}\in\mathbb{C}^{N_{\rm UE}\times N_{\rm UE}}$ and ${\bf A}_{\rm T}\in\mathbb{C}^{N_{\rm AP}\times N_{\rm AP}}$ are discrete Fourier transform matrices, and ${\bf H}_{u,n}^{\rm G,DL}[k]\in\mathbb{C}^{N_{\rm UE}\times NN_{\rm AP}}$ is the sparsified angular-domain channel matrix. Therefore, Eq.~\eqref{eq:pilottrain3} can be further formulated as
\begin{equation}
	\tilde{\bf y}_u^{\rm DL}[k] = {\bf P} {\bf \Psi} {\bf h}_u^{\rm G,DL}[k] + \tilde{\bf n}_u[k],
	\label{eq:pilottrain4}
\end{equation}
where ${\bf \Psi} = {\rm blkdiag}({\bf A}_{\rm T}^{*},\cdots,{\bf A}_{\rm T}^{*})\otimes {\bf A}_{\rm R}\in\mathbb{C}^{NN_{\rm AP}N_{\rm UE}\times NN_{\rm AP}N_{\rm UE}}$ is the dictionary matrix, and ${\bf h}_u^{\rm G,DL}[k] = {\rm vec} ([{\bf H}_{u,1}^{\rm G,DL}[k],\cdots,{\bf H}_{u,N}^{\rm G,DL}[k]])\in\mathbb{C}^{NN_{\rm AP}N_{\rm UE}}$ is the vectorized sparse channel gain vector. 

Similarly, during the uplink channel training phase, as described in Eq.~\eqref{eq:ulprocedure}, $\tau_{\rm AP}$ stacked measurements at the $n$-th AP can be given as
\begin{equation}
	\begin{aligned}
		\tilde{\bf y}_{u,n}^{\rm UL}[k] = {\rm vec}\left( \tilde{\bf Y}_{u,n}^{\rm UL}[k] \right)= {\bf P}^\prime{\bf \Psi}^\prime {\bf h}_{u,n}^{\rm G,UL}[k]+\tilde{\bf n}_n[k],
		\label{eq:pilottrain5}
	\end{aligned}
\end{equation}
where $\tilde{\bf Y}_{u,n}^{\rm UL}[k] \in\mathbb{C}^{\tau_{\rm AP}N_S\times N_{\rm UE}}$ is obtained by collecting ${\bf Y}_{u,n}^{\rm UL}[k]$ in Eq.~\eqref{eq:ulprocedure} for $\tau_{\rm AP}$ time slots, ${\bf h}_{u,n}^{\rm G,UL}[k]={\rm vec}({\bf H}_{u,n}^{\rm G,UL}[k])\in\mathbb{C}^{N_{\rm AP}N_{\rm UE}}$ is the sparsified channel gain vector, ${\bf \Psi}^\prime = {\bf A}_{\rm T}^{*}\otimes {\bf A}_{\rm R}\in\mathbb{C}^{N_{\rm AP}N_{\rm UE} \times N_{\rm AP}N_{\rm UE}}$ is the dictionary matrix, $\tilde{\bf n}_n[k]$ is the vectorized AWGN matrix, and ${\bf P}^\prime = ( {\bf I}_{N_{\rm UE}} \otimes [\tilde{\bf F}_{n}^{1},\cdots,\tilde{\bf F}_{n}^{\tau_{\rm AP}}]^{\rm H}) \in\mathbb{C}^{\tau_{\rm AP} N_S N_{\rm UE}\times N_{\rm AP}N_{\rm UE}}$ is the pilot matrix.

The least squares (LS) method is applicable for solving the channels ${\bf h}_{u}^{\rm G,DL}[k]$ and ${\bf h}_{u,n}^{\rm G,UL}[k]$ in linear measurement problems~\eqref{eq:pilottrain4} and~\eqref{eq:pilottrain5}, under the condition that a sufficient number of measurements are provided. This CSI acquisition method requires $\tau_{\rm UE}\tau_p^{\rm DL}\geq NN_{\rm AP}N_{\rm UE}$ for downlink and $\tau_{\rm AP} N_S\geq UN_{\rm AP}$ for uplink. However, in CF-mMIMO systems, the total number of antennas $NN_{\rm AP}$ may be significantly large, thereby inducing substantial training and computation overhead. Previous works have sought to leverage the inherent sparsity in the angle-domain\cite{dncnn} and consequently formulate the channel acquisition problem as
\begin{equation}
	\begin{aligned}
		{\rm (P1)}\:\underset{{\bf h}_u^{\rm G,DL}[k], 1 \leq k \leq K}{\rm minimize} &  \left\|{\bf h}_u^{\rm G,DL}[k]\right\|_0, \\
		\text { s.t. } & \left\| {\bf P} {\bf \Psi} {\bf h}_u^{\rm G,DL}[k]-\tilde{\bf y}_u^{\rm DL}[k]\right\|_2 \leq \epsilon, \forall k,\\
		& \ {\rm supp}\!\left\{\! {\mathcal S}\! \right\}\!=\!{\rm supp}\!\left\{\! {\mathcal S}_1\! \right\}\!=\!\cdots\!=\!{\rm supp}\!\left\{\! {\mathcal S}_K\! \right\}\!,\notag
	\end{aligned}
	\label{eq:p1}
\end{equation}
which is a structured compressive sensing problem to simultaneously estimate $\{{\bf h}_u^{\rm G,DL} [k]\}_{k=1}^K$ with common sparse support sets $\{\mathcal{S}_k\}_{k=1}^K$. ${\rm (P1)}$ has been extensively investigated and can be resolved iteratively by CS-based algorithms such as orthogonal matching pursuit (OMP) and AMP\cite{dlcemumassiveMIMO,dlcemumassiveMIMO_OMP,FDD,domp}. In FDD mode operations, the downlink CSI, i.e., the solution of ${\rm (P1)}$, still requires to be fed back to APs for multi-user beamforming design. Conventional pre-designed codebook-based CSI feedback can be used to mitigate feedback overhead, but the CSI feedback error can severely degrade the beamforming performance when the dimension of CSI becomes large\cite{dncnn}.

The aforementioned issues have prompted us to explore efficient methods for further minimizing the channel acquisition overhead for both the downlink estimation and feedback procedure for CF-mMIMO systems.

\begin{figure}[t]
	\centering
	\includegraphics[width=0.5\textwidth]{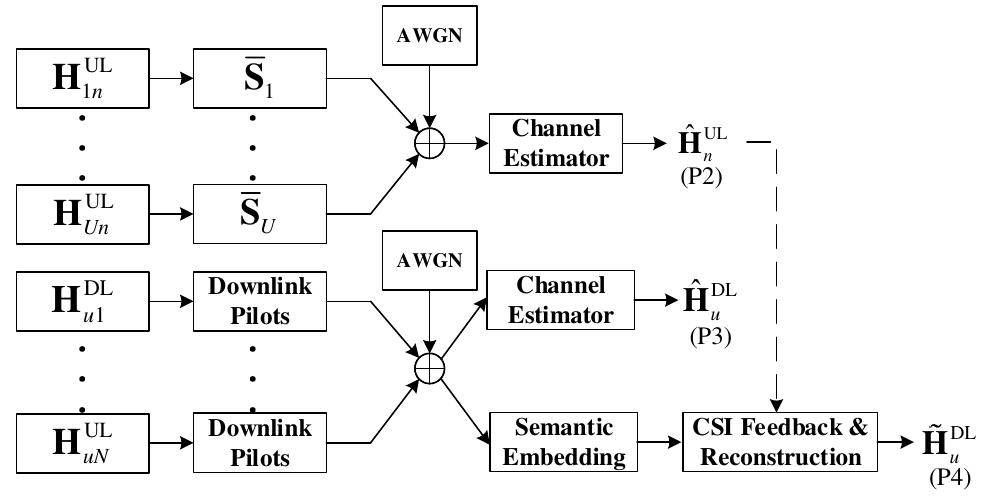}
	\caption{Schematic diagram of the proposed channel acquisition schemes. The received pilots via wireless channels are used for uplink/downlink CSI acquisition, where the uplink/downlink estimation is used for signal detection, while CSI feedback is used for downlink beamforming design.}\label{scheme}
\end{figure}
\subsection{Proposed Uplink Channel Estimation}
Uplink estimation problems can be formulated in a manner similar to ${\rm (P1)}$. Conventional estimation techniques demand orthogonal pilot signals to distinguish CSI from different UEs, leading to a relatively high training overhead. In this section, we introduce a data-driven uplink CSI estimation framework, comprising two fundamental components: (a) the joint design of uplink pilot signals, and (b) the acquisition of CSI from compressed pilots at the AP. Unlike ${\rm (P1)}$, the problem of uplink CSI estimation can be expressed as\footnote{We consider hybrid digital-analog arrays by default, but the same methods can be easily applied to fully digital arrays as special cases.}
\begin{subequations}
	\begin{align}
		{\rm (P2)}\!\!\!\!\underset{\hat{\bf H}_{u,n}^{\rm UL}[k], 1 \leq k \leq K}{\rm minimize} &\ \sum_{k=1}^K \left[ \frac{\Vert \hat{\bf H}_{u,n}^{\rm UL}[k]-{\bf H}_{u,n}^{\rm UL}[k] \Vert_F^2}{\Vert {\bf H}_{u,n}^{\rm UL}[k] \Vert_F^2} \right], \\
		\text { s.t. } &\ \hat{\bf H}_{u,n}^{\rm UL}[k]\!= \notag\\&\ \! f^{\rm UL}\left( {\bf Y}_{u,n}^{\rm UL}[k]\mid {\{ \overline{\bf F}_{n,\tau}^{{\rm RF}}  {\overline{\bf F}_{\tau}^{\rm BB}}[k]\}_{\tau=1}^{\tau^\prime_{\rm AP}}},\overline{\bf S}_u \right), \\
		&\ \sum_{k=1}^K\Vert \overline{\bf F}_{n,\tau}^{\rm RF} \overline{\bf F}_{n,\tau}^{\rm BB}[k] \Vert_F^2 \leq N_S,\notag\\ &\ \tau = 1,\cdots, \tau_{\rm AP}^\prime; \forall n=1\cdots,N,\\ 
		&\ \Vert \overline{\bf F}_{n,\tau}^{\rm RF} \Vert = \frac{1}{\sqrt{N_{\rm AP}}},\forall n=1\cdots,N,\\
		&\ \Vert {\bf S}_u \Vert_F^2 \leq P_{\rm UE}\tau_p^{\rm UL},\forall n=1\cdots,N,
	\end{align}
\end{subequations}
where $P_{\rm UE}$ denotes the transmit power of UE, $f^{\rm UL}(\cdot)$ denotes the proposed mapping from pilots to estimation, as depicted in Fig.~\ref{scheme} and \ref{mixer}, and {$\overline{\bf F}_{n,\tau}^{{\rm RF}}$, ${\overline{\bf F}_{\tau}^{\rm BB}}[k]$ ($\tau = 1,\cdots,\tau_{\rm AP}^\prime$)}, and $\overline{\bf S}_u$ are the considered trainable parameters of $f^{\rm UL}(\cdot)$. We can build the optimization model by means of NN components, where for the $n$-th AP's analog combiner part $\overline{\bf F}_{n,\tau}^{{\rm RF}}\in\mathbb{C}^{N_{\rm AP}\times N_{\rm RF}}$, we define a trainable parameter matrix {in the $\tau$-th slot} ${\boldsymbol{\Theta}_{\tau}^{\rm RF}}$ as
\begin{equation}
	\overline{\bf F}_{n,\tau}^{{\rm RF}}=\frac{1}{\sqrt{N_{\rm AP}}}\left( \cos \left( {\boldsymbol{\Theta}_{\tau}^{\rm RF}} \right)+j\sin \left( {\boldsymbol{\Theta}_{\tau}^{\rm RF}} \right) \right),
	\label{eq:analogbf}
\end{equation}
while for the $n$-th AP's digital combiner part and UE's transmit pilot signals we also define trainable parameter sets ${\overline{\bf F}_{\tau}^{\rm BB}}[k]\in\mathbb{C}^{N_{\rm RF}\times N_S}, \forall k=1\cdots,K$ and $\overline{\bf S}_u\in\mathbb{C}^{N_{\rm UE}\times \tau_p^{\rm UL}}$, respectively. Since it is difficult to directly constrain the power for NN components, we add a normalization operation in feed-forward procedure as
\begin{equation}
	\overline{\bf Y}_{u,n,\tau}^{\rm UL}[k] = \frac{{{\overline{\bf F}_{\tau}^{\rm BB}}}[k]{\overline{\bf F}_{n,\tau}^{{\rm RF}}}}{\left\Vert  {{\overline{\bf F}_{\tau}^{\rm BB}}}[k]{\overline{\bf F}_{n,\tau}^{{\rm RF}}} \right\Vert_F} \left( {\bf H}_{u,n}^{\rm UL}[k]\frac{\overline{\bf S}_u}{\Vert \overline{\bf S}_u \Vert_F} +\overline{\bf N}_n \right).
\end{equation}

Note that in the proposed optimization problem $\rm (P2)$, since the pilot signals $\{\overline{\bf S}_u\}_{u=1}^U$ for different UEs are jointly designed, the orthogonality requirement for different UEs' pilot signals is no longer required to distinguish uplink CSI from different UEs.

\subsection{Proposed Downlink Channel Acquisition from Semantic Embedding}
\begin{figure*}[t]
	\centering
	\includegraphics[width=0.85\textwidth]{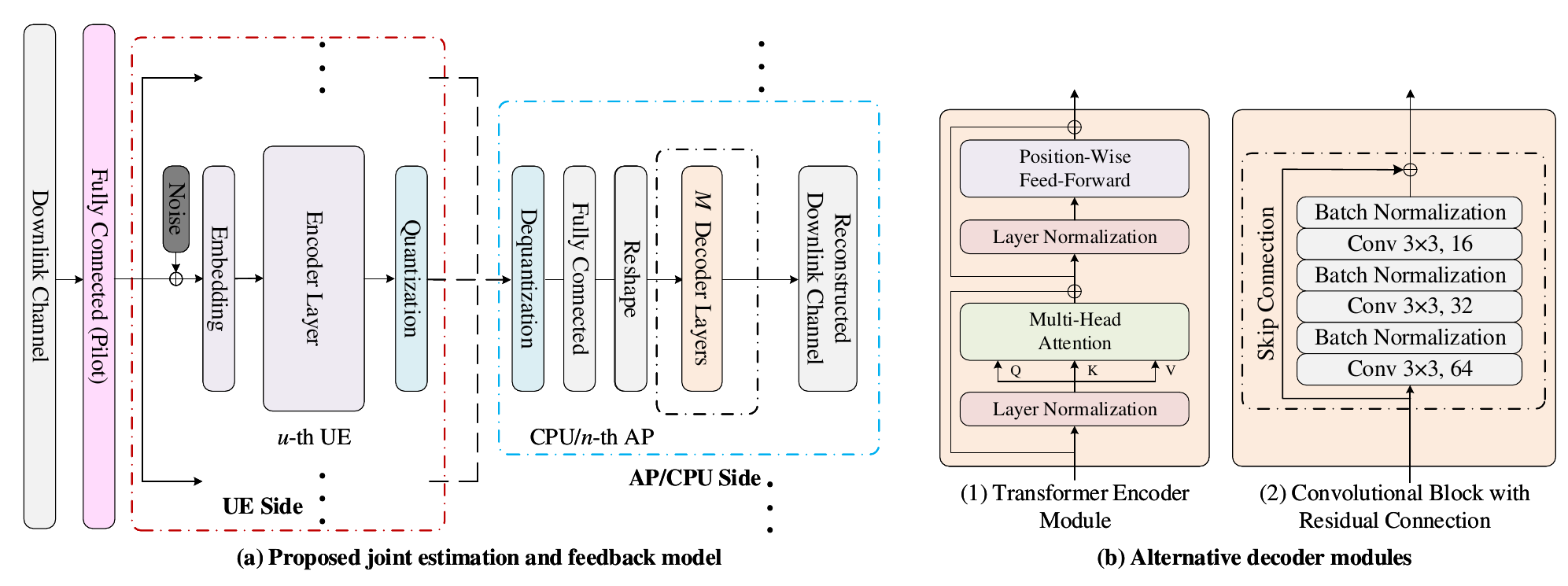}
	\caption{The schematic of (a) proposed Mixer-based feedback model with quantization and (b) alternative decoder modules\cite{attn,csinet}. Models deployed at UE side can share the same parameter setup, and so do the models at AP/CPU side.}\label{fbmodel}
\end{figure*}

As shown in Fig.~\ref{scheme}, downlink CSI needs to be obtained at both AP and UE for designing beamformer and combiner. Consider that the CSI semantic required by beamforming design lies in beamspace information, not necessarily minimizing the normalized mean square error (NMSE) of the channel estimation, we hence adopt the cosine similarity function (CSF) as the evaluation metric in training as
\begin{equation}
	{\rm Loss}({\bf V}_1,{\bf V}_2) = \mathbb{E}\left[ \frac{\vert {\rm vec}\left( {\bf V}_1 \right)^{\rm H} {\rm vec}\left( {\bf V}_2 \right) \vert}{\Vert {\rm vec}( {\bf V}_1 ) \Vert_2  \left\Vert {\rm vec}\left( {\bf V}_2 \right) \right\Vert_2} \right].
\end{equation}
For downlink training at the UE side, the measurement model is shown in Eq.~\eqref{eq:pilottrain2}. Analogous optimization problem can be established to acquire the CSI at the UE through NN components as
\begin{subequations}
	\begin{align}
		{\rm (P3)}\!\!\!\! \underset{\hat{\bf H}_{u}^{\rm DL}[k], 1 \leq k \leq K}{\rm minimize} &\  {\rm Loss} \left( \hat{\bf H}_{u}^{\rm DL},{\bf H}_{u}^{\rm DL} \right) ,\\
		\text { s.t. }\ & \hat{\bf H}_{u}^{\rm DL}[k]= f^{\rm DL}\left( \tilde{\bf y}_u^{\rm DL}[k]\mid { \{ \overline{\bf w}_{u,\tau} \}_{\tau = 1}^{\tau_{\rm UE}^\prime}},\right.\label{eq:dlest}\\
		&\quad\quad\quad\quad\quad\quad\quad \left. {\{ \overline{\tilde{\bf f}}_{\tau^{\prime}} \}_{\tau^\prime=1}^{\tau_p^{\rm DL}}}  \right),\notag\\
		& \Vert \overline{\tilde{\bf f}}_{\tau^{\prime}} \Vert_F^2 \leq P_{\rm AP},\forall n=1\cdots,N,\\ 
		&\Vert \overline{\bf w}_{u,\tau}\Vert_F\leq 1,\ \tau = 1,\cdots,\tau_{\rm UE}^\prime\\
		& \Vert \overline{\bf F}^{\rm RF}_n \Vert = \frac{1}{\sqrt{N_{\rm AP}}},\forall n=1\cdots,N,
	\end{align}
\end{subequations}
where $P_{\rm AP}$ denotes the transmit power of AP, and $f^{\rm DL}(\cdot)$ denotes the proposed mapping from pilot signals to the estimated channel, which is also shown in Fig.~\ref{scheme} and \ref{mixer}. $\hat{\bf H}_{u}^{\rm DL} = [\hat{\bf H}_{u}^{\rm DL}[1],\cdots,\hat{\bf H}_{u}^{\rm DL}[K]]\in\mathbb{C}^{N_{\rm UE}\times KNN_{\rm AP}}$ and ${\bf H}_{u}^{\rm DL} = [{\bf H}_{u}^{\rm DL}[1],\cdots,{\bf H}_{u}^{\rm DL}[K]]\in\mathbb{C}^{N_{\rm UE}\times KNN_{\rm AP}} $ denote the estimated and target channel matrices on all $K$ subcarriers, respectively. 
We can therefore build the optimization problem by NN components as
\begin{equation}
	\begin{aligned}
		\overline{y}_{u,\tau,\tau^\prime}^{\rm DL}[k] =\! \frac{\overline{\bf w}_{u,\tau}^{\rm H}}{\Vert \overline{\bf w}_{u,\tau}^{\rm H} \Vert_F}\left( \frac{\sqrt{P_{\rm AP}}{\bf H}_{u}^{\rm DL}[k]\overline{\tilde{\bf f}}_{\tau^\prime} }{\Vert \overline{\tilde{\bf f}}_{\tau^\prime} \Vert_F}\! +\! \overline{\bf n}_{u,\tau,\tau^\prime}[k] \right),
	\end{aligned}
	\label{eq:learnedmeasurement}
\end{equation}
where $\overline{\bf w}_{u,\tau}$ ($\tau = 1,\cdots,\tau^\prime_{\rm UE}$) and $\overline{\tilde{\bf f}}_{\tau^\prime}$ ($\tau^\prime=1,\cdots,\tau_p^{\rm DL}$) are respectively the trainable version of combining vectors and beamforming vectors in Eq.~\eqref{eq:pilottrain1}.

The downlink channels estimated at the UE need to be fed back to APs for beamforming design. Given the potentially high dimensionality of the estimated CSI matrix, we propose to feedback the quantized semantic embedding vectors (SEV) of the estimated downlink channels, as opposed to compressing the entire estimated channel directly in conventional schemes\cite{csinet,feedbackdl,selfinfofb,MCfb}. This processing can be formulated as
\begin{subequations}
	\begin{align}
		{\rm (P4)}\!\! \underset{\tilde{\bf H}_{u}^{\rm DL}[k], 1 \leq k \leq K}{\rm minimize} &\:  {\rm Loss}\left( \tilde{\bf H}_{u}^{\rm DL}, \tilde{\bf w}_u^{\rm H}{\bf H}_{u}^{\rm DL} \right),\\
		\text { s.t. }\ & \tilde{\bf H}_{u}^{\rm DL} = f_{\rm rec}\left( {\bf \mathcal{X}}_u \right),\label{eq:recmodel}\\
		&{\bf \mathcal{X}}_u = q\left(f_{\rm emb}\left({\overline{\bf Y}_u^{\rm DL}[1],\cdots,\overline{\bf Y}_u^{\rm DL}[K]}\right)\right),\\
		& \Vert \overline{\tilde{\bf f}}_{\tau^{\prime}} \Vert_F^2 \leq P_{\rm AP},\forall n=1\cdots,N,\\ 
		&\Vert \overline{\bf w}_{u,\tau}\Vert_F\leq 1,\ \tau = 1,\cdots,\tau_{\rm UE}^\prime\\
		& \Vert \overline{\bf F}^{\rm RF}_n \Vert = \frac{1}{\sqrt{N_{\rm AP}}},\forall n=1\cdots,N,
	\end{align}
\end{subequations}
where $f_{\rm emb}(\cdot)$ and $f_{\rm rec}(\cdot)$ represent the proposed channel SEV and reconstruction modules, respectively, as depicted in Fig.~\ref{fbmodel}, $q(\cdot)$ is the quantization operation, $\overline{\bf Y}_u^{\rm DL}[k]$ is the overall measurement matrix similar to $\tilde{\bf Y}_u^{\rm DL}[k]$ in Eq.~\eqref{eq:pilottrain2} composed of $\overline{\tilde{y}}_{u,\tau,\tau^\prime}^{\rm DL}[k]$ in Eq.~\eqref{eq:learnedmeasurement}, and ${\bf \mathcal{X}}_u\in\mathbb{Z}^{B}$ is the SEV. $\tilde{\bf w}_u\in\mathbb{C}^{N_{\rm UE}}$ is the optimal single stream UE combining vector, which can be obtained from the eigenvector corresponding to the maximum eigenvalue of ${\bf H}^{\rm DL}_u[k]({\bf H}^{\rm DL}_u[k])^{\rm H}$.

To further reduce the feedback overhead $B$, angle-domain reciprocity of uplink/downlink channels can be utilized for APs to reconstruct partial CSI from shorter feedback sequences with the help of estimated uplink CSI, where we only need to modify $\rm (P4)$ in Eq.~\eqref{eq:recmodel} as
\begin{equation}
	\tilde{\bf H}_{u}^{\rm DL} = f_{\rm rec}^\prime\left( {\bf \mathcal{X}}_u ; \hat{\bf H}_{n}^{\rm UL} \right).
\end{equation}

\subsection{Network Backbone Structure}

Previous approaches\cite{fb3GPP,fddcf1} have focused on feeding the feature of interest of the estimated channel matrix (e.g., the dominant path\cite{fddcf1}) to the APs. However, the use of NN tools brings the possibility of compressing the CSI with a higher information density, resulting in the creation of so-called channel semantic vectors.

\begin{figure*}[tbp]
	\centering
	\includegraphics[width=0.85\textwidth]{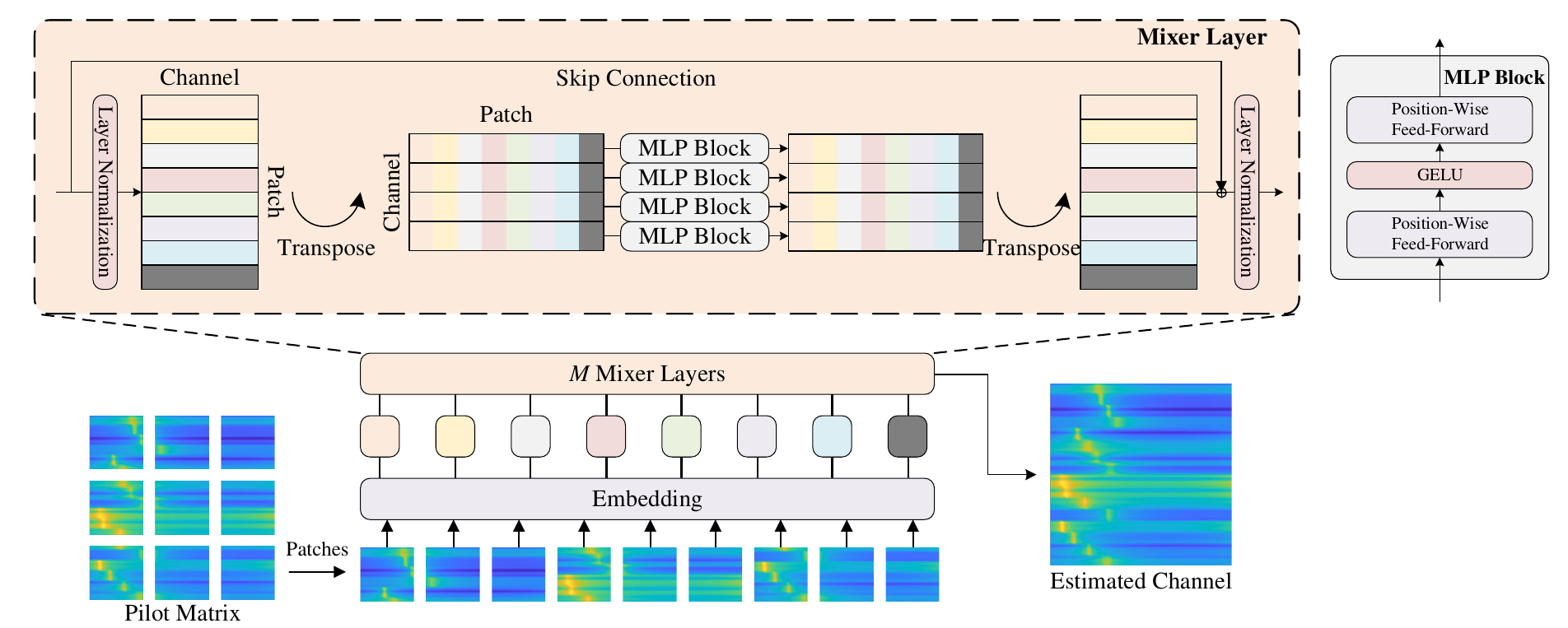}
	\caption{The schematic diagram shows the backbone NN model featuring the innovative MLP-Mixer building block and its corresponding MLP block. The pilot matrix received is divided into segments and integrated by linear embedding for efficient Mixer operations.}\label{mixer}
\end{figure*}

We therefore seek models with customized mechanisms for faster inference speed and more accurate fine-grained features\cite{attn,mlpmixer}. MLP-Mixer (or Mixer for convenience) is one of these state-of-the-art model, as shown in Fig.~\ref{mixer}, which not only has global perception, but also occupies less memory than CNN- and Transformer-based models\cite{attn}. The MLP blocks adopted in the Mixer layers is illustrated in Fig.~\ref{mixer}, which is modified from conventional MLP-Mixer. 

As shown in Fig.~\ref{scheme} and Fig.~\ref{fbmodel}, instead of compressing the CSI matrix estimated in Eq.~\eqref{eq:dlest}, we choose to create the semantic vector directly from the noisy received pilot matrix ${{\bf Y}_u^{\rm DL}} = [ {\bf y}_u^{\rm DL}[1],{\bf y}_u^{\rm DL}[2],\cdots,{\bf y}_u^{\rm DL}[K] ]\in\mathbb{C}^{\tau_{\rm UE}\tau_p^{\rm DL}\times K}$. Specifically, we select $N_{\rm W}$ and $N_{\rm H}$ that can be evenly divided by $\tilde{T}=\tau_{\rm UE}\tau_p^{\rm DL}$ and $K$, respectively, and split ${\bf Y}_u^{\rm DL}$ into $N_{\rm patch}={\tilde T}K/N_{\rm W}N_{\rm H}$ patches with size $N_{\rm W}\times N_{\rm H}$. Stacking the real and the imaginary parts on a separate dimension for input embedding operation, we have
\begin{equation}
	{\bf X}^{(0)}_{(i-1)*W+j,k} = \sum_{i^\prime=1}^{W}\sum_{j^\prime=1}^{H}\sum_{c=1}^{2}  \left( {\bf \Xi}^{(k)} \odot \tilde{\bf Y}_{\rm patch}^{(i,j)} \right)_{i^\prime,j^\prime,c},
\end{equation}
where $\tilde{\bf Y}^{(i,j)}_{\rm patch}\in\mathbb{R}^{N_{\rm W}\times N_{\rm H} \times 2}$ is the $(i, j)$-th non-overlapping patch cropped continuously from a 3D matrix (i.e., tensor), and the third dimension of the tensor represents the stacked real and imaginary parts of ${\bf Y}_u^{\rm DL}$.
${\bf \Xi}^{(k)}\in\mathbb{R}^{N_{\rm W}\times N_{\rm H} \times 2}$ denotes the $k$-th ($k=1,\cdots, N_{\rm emb}$) convolutional kernel, and ${\bf X}^{(0)}\in\mathbb{R}^{N_{\rm patch}\times N_{\rm emb}}$ is the input embedding. ${\bf X}^{(0)}$ is then sent to Mixer layers and follows the calculation procedure as
\begin{equation}
	\begin{aligned}
		\tilde{\bf X}^{(m)}_{:,k}&={\bf X}^{(m)}_{:,k}+{\bf W}^{(m)}_2 \sigma\left( {\bf W}^{(m)}_1 {\rm LN}({\bf X}^{(m-1)}_{:,k}) \right)\\
		{\bf X}^{(m)}_{n,:}&=\tilde{\bf X}^{(m)}_{n,:}+ \sigma\left(  {\rm LN}(\tilde{\bf X}^{(m)}_{n,:}){\bf W}^{(m)}_3 \right){\bf W}^{(m)}_4,
	\end{aligned}
\end{equation}
where ${\bf W}^{(m)}_1\in\mathbb{R}^{N_{\rm tok}\times N_{\rm patch}}$, ${\bf W}^{(m)}_2\in\mathbb{R}^{N_{\rm patch}\times N_{\rm tok}}$ ,${\bf W}^{(m)}_3\in\mathbb{R}^{N_{\rm emb}\times N_{\rm hid}}$, ${\bf W}^{(m)}_4\in\mathbb{R}^{N_{\rm hid}\times N_{\rm emb}}$ denote the linear transformation matrices at the $m$-th Mixer layer ($m=1,\cdots,M_{\rm enc}$), ${\rm LN}(\cdot)$ denotes the layer normalization (LN) operation, and $\sigma(\cdot)$ is the activation function. $\tilde{\bf X}^{(m)}\in\mathbb{R}^{N_{\rm tok}\times N_{\rm emb}}$ denotes the intermediate variable in Mixer block. After vectorization, dimensionality reduction, and quantization, we are able to obtain channel semantic vector $\mathcal{X}_u\in\mathbb{Z}^{B}$ for the $u$-th UE that implicitly characterize CSI.

The quantization module $q(\cdot)$ significantly reduces the feedback overhead, as the compressed channel semantic vectors can be represented with only $B$ binarized numbers rather than float numbers. In this work, we assume that the feedback of these compressed semantic vectors is perfect and can be accurately obtained at the APs.

After receiving the feedback semantic vectors $\{\mathcal{X}_u\}_{u=1}^U$, APs will extract channel matrix from such vectors with high information density. Similarly, the decoding part can be achieved with another $M_{\rm dec}$ Mixer layers, as shown in Fig.~\ref{fbmodel}(a). The proposed scheme can be divided into pilot training, channel SEV, and CSI reconstruction modules, as we have mostly discussed before. It is worth noting that we have taken into account the difference in computation resources between UEs and APs, setting $M_{\rm enc} = 1$ for UEs while utilizing more layers $M_{\rm dec}=4$, kernel size, and number of hidden channels for reconstruction model at APs.

Several alternatives are considered as shown in Fig.~\ref{fbmodel}(b), which are typically deployed at APs or CPUs to relax the complexity considerations in practice. For demonstration purposes, we provide practical parameters that are still feasible for use in performance comparison.

\section{Model-Driven Deep-Unfolding Beamforming under Imperfect CSI}
\subsection{Preliminaries}
Asymptotically orthogonal channels in mMIMO systems make it possible for linear beamforming schemes to achieve near-optimal spectral efficiency with much lower computational complexity\cite{asyortho,asyortho2}. However, linear approaches such as ZF still suffers from severe performance deterioration in the case of ill-conditioned channel matrix sensitive to numerical disturbance, as well as computational complexity of high-dimensional matrix inversion problems. Therefore, an optimal linear beamforming named as linear minimum mean square error (LMMSE) is considered to mitigate the performance loss as
\begin{equation}
	\tilde{\bf F}_{\rm lmmse}[k]\!=\!\left(\!\tilde{\bf H}^{\rm DL}[k] \right)^{\rm H}\!\!\left(\!\!\left(\tilde{\bf H}^{\rm DL}[k]\left( \tilde{\bf H}^{\rm DL}[k]\! \right)^{\rm H}\!\!\!\!+\!{\bf C}[k]\!\right) \!\!+\!\sigma_{n}^2 \mathbf{I}\right)^{-1}\!\!\!\! \!\!,
\end{equation}
where ${\bf C}[k]=\mathbb{E}[{\bf e}_k{\bf e}_k^{\rm H}]$ denotes the auto-correlation matrix of the CSI reconstruction error ${\bf e}_k$ on the $k$-th subcarriers, and $\tilde{\bf H}^{\rm DL}[k]\in\mathbb{C}^{UN_{\rm UE}\times NN_{\rm AP}}$ represents the reconstructed channel matrix. We can simplify the beamformer in a regularized form as 
\begin{equation}
	\tilde{\bf F}_{\rm rzf}[k]\!=\!\left( \tilde{\bf H}^{\rm DL}[k] \right)^{\rm H}\!\!\left(\!\tilde{\bf H}^{\rm DL}[k]\left( \tilde{\bf H}^{\rm DL}[k] \right)^{\rm H}\!\!\!+\!{\rm diag}({\boldsymbol{\alpha}})\tilde{\bf C}[k]\right) ^{-1}\!,
\end{equation}
where we use $\tilde{\bf C}[k]\in\mathbb{C}^{UN_{\rm UE}\times UN_{\rm UE} }$ to denote the regularization matrix, and ${\boldsymbol{\alpha}}=[\alpha_1,\cdots,\alpha_{UN_{\rm UE}}]^{\rm H}$ denotes the scaling factor vector. When we have independent identically distributed (i.i.d.) noise and reconstruction error, the regularization term and scaling factor vector degrade to an identical matrix and a scalar, respectively, and the closed-form solution of asymptotically optimal $\alpha^\star$ 
under some additional assumptions can be given as
\begin{align}
	{\bar \alpha}^\star = \left( \frac{\sigma^2_n+\varepsilon^2}{1-\varepsilon^2}\right)\frac{U}{NN_{\rm AP}},
	\label{eq:optimalrfac}
\end{align}
which can be proved under several strong assumptions~\cite{randmat}, including
\begin{itemize}
	\item The considered scenario is a conventional single BS broadcast model.
	\item The number of UEs $U$, the number of APs $N$ and the number of antennas at APs $N_{\rm AP}$ are assumed to be very large for asymptotic analysis.
	\item Channel matrix is assumed to be Gaussian i.i.d. in a homogeneous networks without antenna correlation.
	\item Estimation and feedback errors are assumed to be identical for all UEs, and the distortion $\varepsilon$ can be defined~\cite{randmat} by
	\begin{equation}
		\hat{\bf h} = {\sqrt{1-\varepsilon^2}}{\bf h}+\varepsilon {\bf n}.
	\end{equation}
\end{itemize}

\begin{figure}[tbp]
	\centering
	\includegraphics[width=0.35\textwidth]{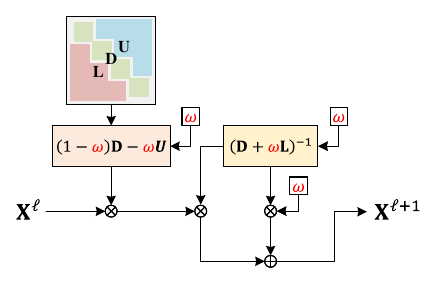}
	\caption{A single iteration process of SOR method. We still use matrix inversion here for simplicity, while inversion operation can be avoided by using Eq.~\eqref{eq:soriter2}.}\label{sorsingle}
\end{figure}

In CF-mMIMO systems, however, the aforementioned assumptions no longer hold, and the optimal solution $\alpha^\star$ for maximizing the sum-rate can hardly be given in analytical form. In order to jointly optimize the regularization factor to combat imperfect CSI feedback and accelerate the convergence process, we adopt an iterative method named successive over-relaxation, and propose a deep-unfolding beamforming method based on it. We define $\tilde{\bf R} = \tilde{\bf H}^{\rm DL}( \tilde{\bf H}^{\rm DL} )^{\rm H}+{\rm diag}(\boldsymbol{\alpha}) {\bf C}$ as the target matrix to be inversed (ignoring subcarrier index $k$ for brevity), and it is obvious that $\tilde{\bf R}\in\mathbb{C}^{UN_{\rm UE}\times UN_{\rm UE}}$ is a diagonally dominant matrix. We firstly consider Gauss-Seidel method, which is a special case of SOR method. Specifically, we define
\begin{align}
	\tilde{\bf R} = {\bf D}_R+{\bf L}_R+{\bf U}_R,
	\label{definedecomp}
\end{align}
where ${\bf D}_R$, ${\bf L}_R$, and ${\bf U}_R$ are respectively the diagonal part, the strictly lower triangular part, and the strictly upper triangular part of $\tilde{\bf R}$, as shown in Fig.~\ref{sorsingle}. Thus, Gauss-Seidel iterative solution for linear equations $\tilde{\bf R}{\bf x} = {\bf b}$ can be expressed as
\begin{align}
	{\bf b} = \left({\bf D}_R+{\bf L}_R\right){\bf x}^{(\ell+1)}+{\bf U}_R{\bf x}^{(\ell)},
\end{align}
where ${\bf x}^{(\ell)}$ denotes the solution of the aforementioned linear equations in the $\ell$-th iteration. The iteration process can be further written as 
\begin{align}
	{\bf x}^{(\ell+1)} = -\left({\bf D}_R+{\bf L}_R\right)^{-1}{\bf U}_R{\bf x}^{(\ell) }+\left({\bf D}_R+{\bf L}_R\right)^{-1}{\bf b}.
\end{align}
In each iteration, the update direction is expressed as ${\bf x}^{(\ell+1)} - {\bf x}^{(\ell)}$, and if we take this direction as a weighted correction, the iteration can be written as
\begin{align}
	{\bf \tilde x}^{(\ell+1)} = {\bf x}^{(\ell)} + \omega \left( {\bf x}^{(\ell+1)} - {\bf x}^{(\ell)} \right),
\end{align}
where $\omega$ is the convergence factor. Given the update process, the iterative update can be given respectively as
\begin{align}
	{\bf x}^{(\ell+1)} = & \left({\bf D}_R+\omega{\bf L}_R\right)^{-1}\left((1-\omega ){\bf D}_R -\omega {\bf U}_R\right){\bf x}^{(\ell)}\notag\\&+\omega \left({\bf D}_R+\omega {\bf L}_R\right)^{-1}{\bf b},
	\label{sormat}
\end{align}
or
\begin{align}
	{x}_{i}^{(\ell+1)} = {x}_{i}^{(\ell)} + \frac{\omega}{r_{i,i}}\left(b_i -\sum_{j=1}^{i-1}r_{i,j}{x}_{j}^{(\ell+1)}- \sum_{j=i+1}^{U}r_{i,j}{x}_{j}^{(\ell)}\right),
	\label{eq:soriter}
\end{align}
where $r_{i,j}$ denotes the element in the $i$-th row and the $j$-th column of matrix $\tilde{\bf R}$.
Apparently, SOR method is compatible for $N$ parallel linear equations as ${\bf X}_R^\ell = [{\bf x}_1^\ell,\dots,{\bf x}_N^\ell]$ and ${\bf B}_R = [{\bf b}_1,\dots,{\bf b}_N]$.
By assigning ${\bf B}_R = {\bf I}$, the aforementioned SOR method is capable of solving the matrix inversion problem in an iterative manner.

Introducing SOR method can effectively accelerate the convergence, while the optimal relaxation factor $\omega$ and regularization factor $\alpha$ are hard to obtain. Inspired by the model driven deep unfolding methods\cite{modeldri,mxsjsac}, we have proposed a learning-based unfolding method for beamforming design. As shown in Fig. \ref{sorsingle}, the unfolded model takes one single iteration process as one layer of the neural network. By integrating the trainable parameters into the unfolded iterations of conventional SOR method, the model can adaptively learn and optimize the parameter from data samples.

Assume that the transfer function of the $\ell$-th SOR iteration is $f_\ell(\cdot)$, then the inversion operation is approximated as 
\begin{align}
	\left(\tilde{\bf R}^{(\ell)}\right)^{-1} &= f_\ell\left(\left(\tilde{\bf R}^{(\ell-1)}\right)^{-1}\right)\notag\\&=f_\ell\left(f_{\ell-1}\left(\cdots f_{1} \left(\left(\tilde{\bf R}^{(0)}\right)^{-1}\right)\right)\right),
	\label{eq:soriter2}
\end{align}
where $(\tilde{\bf R}^{(\ell)})^{-1}, \ell>0$ denotes the output of the $\ell$-th iteration. The initial value of the iteration is selected as $(\tilde{\bf R}^{(0)})^{-1} = {\rm diag}(1/{\rm diag}(\tilde{\bf R}))$, where $\tilde{\bf R} = \tilde{\bf H}^{\rm DL}( \tilde{\bf H}^{\rm DL} )^{\rm H}+{\rm diag}(\boldsymbol{\alpha}) {\bf C}$, and $\tilde{\bf H}^{\rm DL}$ is the reconstructed channel matrix from the feedback model. 
\subsection{Fully-Digital Array Architecture}

For the centralized processing paradigm, signals are encoded at the CPU before distributed to APs. Assume that all APs serve all UEs, the linear beamforming matrix on the $k$-th subcarrier can be further obtained as
\begin{align}
	{\bf \tilde F}^{\rm cen}_{\rm rzf}[k] = \beta^{\rm cen}\left(\tilde{\bf H}^{\rm DL}[k]\right)^{\rm H}\left(\tilde{\bf R}^{(L_{\rm max})}[k]\right)^{-1},
	\label{eqprecrzf}
\end{align}
where $L_{\rm max}$ denotes the maximum number of iterations, and $(\tilde{\bf R}^{(L_{\rm max})}[k])^{-1}$ denotes the SOR-based output after $L_{\rm max}$ iterations according to Eq.~\eqref{eq:soriter}. To ensure the interference mitigation of RZF-based method, the power normalization factor $\beta^{\rm cen}$ is given as
\begin{equation}
	\beta^{\rm cen} = \underset{1\leq n\leq N}{\rm min}\ {1}/{\left\Vert\left( {\bf \tilde F}^{\rm cen}_{\rm rzf}[k] \right)_{(n-1)N_{\rm AP}+1:nN_{\rm AP},:}\right\Vert_F}.
	\label{eq:normcen}
\end{equation}
For the distributed processing paradigm, signals are distributed to APs and encoded locally with partial CSI $\tilde{\bf H}_n^{\rm DL}=\tilde{\bf H}^{\rm DL}_{:,(n-1)N_{\rm AP}+1:nN_{\rm AP}}$ at each AP. The received $N_{\rm S}\leq N_{\rm AP}$ data streams at the $n$-th AP is scheduled by $\tilde{\bf D}_n$, thus beamforming matrix can be designed by
\begin{align}
	{\bf \tilde F}_{\rm rzf}^{{\rm dis},n}[k] = \beta^{\rm dis}_n \left(\tilde{\bf D}_n^{\rm T}{\bf \hat H}^{\rm DL}_{n}[k]\right)^{\rm H}\left(\tilde{\bf R}^{(L_{\rm max})}_n[k]\right)^{-1},
	\label{eqprecrzfp}
\end{align}
where ${\bf \hat R}_n^{(L_{\rm max})}$ denotes the SOR-based inversion result for ${\bf \hat R}_n = \tilde{\bf D}_n^{\rm T}\tilde{\bf H}^{\rm DL}_{:,(n-1)N_{\rm AP}+1:nN_{\rm AP}}( \tilde{\bf H}^{\rm DL}_{:,(n-1)N_{\rm AP}+1:nN_{\rm AP}} )^{\rm H}\tilde{\bf D}_n + \boldsymbol{\alpha} {\bf C}$ after $L_{\rm max}$ iterations. Normalization factor is given by
\begin{equation}
	\beta_n^{\rm dis}={1}/{\Vert {\bf \tilde F}_{\rm rzf}^{{\rm dis},n}[k] \Vert_F}.
	\label{eq:normdis}
\end{equation}

\SetAlgoNoLine
\SetAlFnt{\small}
\SetAlCapFnt{\normalsize}
\SetAlCapNameFnt{\normalsize}
\begin{algorithm}[!t]
	\caption{Proposed Fully-Digital Beamforming Design}\label{fdsumrate}
	\begin{algorithmic}[1]
		\REQUIRE The acquired downlink CSI matrix $\tilde{\bf H}^{\rm DL}[k]$ or partial CSI $\tilde{\bf H}_n^{\rm DL}[k]$ at the APs. Initial value $\omega = 1$ and ${\bf \alpha} = 0$. Maximum iteration $L_{\rm max}$.
		\ENSURE The beamforming matrix $\tilde{\bf F}^{\rm cen}_{\rm rzf}$ or $\tilde{\bf F}^{{\rm dis},n}_{\rm rzf}$.
		\STATE ${\forall k,u,n}$: Set iteration index $\ell$ to 1. Initialize $\omega$ and $\alpha$.
		\IF{Centralized Processing Paradigm}
		\FOR {$ k=1 $ to $K$}
		\STATE Calculate $\tilde{\bf R}^{(0)}[k] = \tilde{\bf H}^{\rm DL}[k](\tilde{\bf H}^{\rm DL}[k])^{\rm H}+{\rm diag}({\boldsymbol{\alpha}}){\bf C}$
		\STATE Calculate SOR iteration using Eq.~\eqref{eq:soriter} for $L_{\rm max}$ times.
		\STATE Calculate beamforming matrix using Eq.~\eqref{eqprecrzf} and Eq.~\eqref{eq:normcen}
		\ENDFOR
		\ELSE[Distributed Processing Paradigm]
		\FOR {$ k=1 $ to $K$}
		\STATE Calculate $\tilde{\bf R}_n^{(0)}[k] = \tilde{\bf H}^{\rm DL}_n[k](\tilde{\bf H}^{\rm DL}_n[k])^{\rm H}+{\rm diag}({\boldsymbol{\alpha}}){\bf C}$
		\STATE Calculate SOR iteration using Eq.~\eqref{eq:soriter} for $L_{\rm max}$ times.
		\STATE Calculate beamforming matrix using Eq.~\eqref{eqprecrzfp} and Eq.~\eqref{eq:normdis}
		
		\ENDFOR
		\ENDIF
		\STATE Calculate loss function according to Eq.~\eqref{eqloss} and perform back propagation for parameters ($\boldsymbol{\omega}$, $\boldsymbol{\alpha}$, $\bf C$) till convergence.
	\end{algorithmic}
\end{algorithm}

In order to optimize the integrated parameters such as $\boldsymbol{\alpha}$, $\bf C$, and $\omega$ to improve the SE performance, the loss function of the model-driven beamforming module is given as
\begin{align}
	R(\omega, {\boldsymbol{\alpha}}, {\bf C}) = \frac{\tau}{\tau+\tau_p}\sum_{k=1}^{K}\sum_{u=1}^{U} \log_2\left(1 + { \gamma}_{u,k}\right),
	\label{eq:SE}
\end{align}
where $\tau$ is the duration of payload transmission, $\tau_p$ is the duration of pilot training (e.g., $\tau_p^{\rm UL}$ and $\tau_p^{\rm DL}$), and $\gamma_{u,k}$ is the signal to noise and interference ratio (SINR). Specifically, for centralized beamforming processing paradigm, SINR is given by
\begin{equation}
	{\gamma}_{u,k}=\frac{\displaystyle\left\|{\bf w}_u^{\rm H}[k] \sum_{n=1}^{N} {\bf H}_{u,n}^{\rm DL}[k]{\bf f}_{u,n}^{\rm cen}[k] \right\|^2}{\sigma_n^2 + \displaystyle  \left\|{\bf w}_u^{\rm H}[k] \sum_{n=1}^N\sum_{i\neq u}^U {\bf H}_{u,n}^{\rm DL}[k] {\bf f}_{n,i}^{\rm cen}[k] \right\|^2},
	\label{sinr}
\end{equation}
and for distributed systems, SINR is given by
\begin{align}
	{ \gamma}_{u,k} =\frac{\displaystyle \sum_{n=1}^N \left\Vert {\bf w}_u^{\rm H}[k] {\bf H}_{u,n}^{\rm DL}[k] {\bf f}_n^{\rm dis}[k] \right\Vert^2}{\displaystyle \sigma_n^2 + \sum_{n=1}^N\displaystyle  \left\|{\bf w}_u^{\rm H}[k] \sum_{i\neq u}^U {\bf H}_{u,n}^{\rm DL}[k] {\bf f}_{n}^{\rm dis}[k] \right\Vert^2},
	\label{sinr2}
\end{align}
Eq.~\eqref{eq:SE} can be further rewritten as loss function to be optimized by gradient methods, which is supported by mainstream learning frameworks as
\begin{align}
	{J}\left(\omega,{\boldsymbol{\alpha}},{\bf C} \right) = -{\hat R}(\omega, \boldsymbol{\alpha},{\bf C}),
	\label{eqloss}
\end{align}
where in ${\hat R}(\omega, \boldsymbol{\alpha})$ we use the reconstructed channel ${\bf \hat H}^{\rm DL}_{u,n}[k]$ for sum-rate calculation. The overall beamforming process is summarized in Algorithm \ref{fdsumrate}.

\subsection{Extended to Hybrid Analog-Digital Architecture}
\SetAlgoNoLine
\SetAlFnt{\small}
\SetAlCapFnt{\normalsize}
\SetAlCapNameFnt{\normalsize}
\begin{algorithm}[!t]
	\caption{Proposed Hybrid Analog-Digital Beamforming}\label{hybsumrate}
	\begin{algorithmic}[1]
		\REQUIRE The acquired CSI matrix $\tilde{\bf H}^{\rm DL}[k]$ or partial CSI $\tilde{\bf H}_n^{\rm DL}[k]$ at the APs. Initial value $\omega = 1$ and ${\bf \alpha} = 0$. Maximum iteration $L_{\rm max}$.
		\ENSURE The beamforming matrix ${\bf F}^{\rm cen}_{\rm RF}$ and ${\bf F}^{\rm cen}_{\rm BB}[k]$, or ${\bf F}^{{\rm dis},n}_{\rm RF}$ and ${\bf F}^{{\rm dis},n}_{\rm BB}[k]$ $\forall\ 1\leq n \leq N,\: 1\leq k \leq K$.
		\STATE ${\forall k,u,n}$: Set iteration index $i$ to 1. Initialize $\omega=\omega_0$ and $\alpha=\alpha_0$.
		\IF{Centralized Processing Paradigm}
		\STATE Calculate the analog beamformer $ {\bf F}_{\rm RF} $ using Eq.~\eqref{eq:cenhyb}.
		\FOR {$ k=1 $ to $K$}
		\STATE Calculate equivalent baseband channel ${\bf H}_{\rm eq}[k]=\tilde{\bf H}^{\rm DL}[k]{\bf F}_{\rm RF}^{\rm cen}$
		\STATE Calculate digital beamformer using Eq.~\eqref{eq:rzfhyb} 
		\ENDFOR
		\ELSE[Distributed Processing Paradigm]
		\STATE Calculate the analog beamformer $ {\bf F}_{\rm RF} $ using Eq.~\eqref{eq:dishyb}.
		\FOR {$ k=1 $ to $K$}
		\STATE Calculate equivalent baseband channel ${\bf H}_{\rm eq}^n[k]=\tilde{\bf H}^{\rm DL}_n[k]{\bf F}_{\rm RF}^{n}$
		\STATE Calculate digital beamformer using Eq.~\eqref{eq:rzfhyb2} 
		\ENDFOR
		\ENDIF
		\STATE Calculate loss function according to Eq. \eqref{eqloss} and perform back propagation for parameters ($\boldsymbol{\omega}$, $\boldsymbol{\alpha}$, $\bf C$) till convergence.
	\end{algorithmic}
\end{algorithm}

Hybrid beamforming/combining architectures are cost-effective alternatives to achieve mMIMO deployment, while optimal design of such beamformers still remains to be further investigated. Most existing works formulate hybrid beamforming as matrix factorization problems\cite{yu2016hybrid}, which assume full knowledge of CSI at transmitter for optimal singular value decomposition or block diagonalization fully-digital beamforming design. 

In this section, due to the complex constraints inherent in hybrid beamforming, imperfect CSI reconstructed at the APs, and required 
low complexity, hybrid beamformers are designed in a two-step manner. We firstly design the analog beamforming part with unit modulus constraint using an equal gain transmission (EGT) to harvest the array gain\cite{dong2014pzf} as
\begin{equation}
	\left( {\bf F}_{\rm RF}^{\rm cen} \right)_{\{i,j\}} =\frac{\left( \displaystyle\sum_{k=1}^K \tilde{\bf H}^{\rm DL}[k] \right)^{\rm H}_{\{j,i\}}}{\left\vert \left( \displaystyle\sum_{k=1}^K \tilde{\bf H}^{\rm DL}[k] \right)_{\{j,i\}}^{\rm H} \right\vert},
	\label{eq:cenhyb}
\end{equation}
and the digital part is designed at CPU using RZF method as
\begin{equation}
	{\bf F}_{\rm BB}^{\rm cen} = \left( \beta^{\rm cen} \right)^\prime \left(\tilde{\bf R}^{(L_{\rm max})}[k]\right)^{-1}\left({\bf H}_{\rm eq}[k]\right)^{\rm H},
	\label{eq:rzfhyb}
\end{equation}
where ${\bf H}_{\rm eq}[k]=\tilde{\bf H}^{\rm DL}[k]{\bf F}_{\rm RF}^{\rm cen}\in\mathbb{C}^{U\times U}$ denotes the equivalent baseband channel matrix, and $( \beta^{\rm cen} )^\prime={\rm min}\: (1/\Vert ({\bf F}_{\rm RF}^{\rm cen})_{(n-1)N_{\rm AP}+1:nN_{\rm AP},:} {\bf F}_{\rm BB}^{\rm cen} \Vert_F)$ is the normalization factor. However, for distributed processing paradigm, the analog part is designed based on the scheduling matrix as 
\begin{equation}
	\left( {\bf F}_{\rm RF}^{{\rm dis},n} \right)_{\{i,j\}} =\frac{\left( \displaystyle\sum_{k=1}^K \tilde{\bf D}_n\tilde{\bf H}^{\rm DL}_n[k] \right)_{\{j,i\}}^{\rm H}}{\left\vert \left( \displaystyle\sum_{k=1}^K \tilde{\bf D}_n\tilde{\bf H}^{\rm DL}_n[k] \right)_{\{j,i\}}^{\rm H} \right\vert},
	\label{eq:dishyb}
\end{equation}
and the digital part is designed locally as
\begin{equation}
	{\bf F}_{\rm BB}^{{\rm dis},n} = \left( \beta^{\rm dis}_n \right)^\prime\left(\tilde{\bf R}^{(L_{\rm max})}[k]\right)^{-1}\left({\bf H}_{\rm eq}^n[k]\right)^{\rm H}\tilde{\bf D}_n,
	\label{eq:rzfhyb2}
\end{equation}
where ${\bf H}_{\rm eq}^n[k]=\tilde{\bf H}^{\rm DL}_n[k]{\bf F}_{\rm RF}^{n}\in\mathbb{C}^{U\times N_{\rm S}}$, and $\left( \beta^{\rm dis}_n \right)^\prime = 1/\Vert{\bf F}_{\rm RF}^{{\rm dis},n}{\bf F}_{\rm BB}^{{\rm dis},n}\Vert_F$. The overall beamforming procedure is summarized in Algorithm.~\ref{hybsumrate}.

\section{Complexity Analysis}
\renewcommand\arraystretch{1.25}
\begin{table*}[!t]
	\centering
	\caption{Computational Complexity of Downlink Channel Acquisition Algorithms}
	\vspace{-2mm}
	\label{tab:complexity}
	\begin{tabular}{c|c}
		\hline\hline
		Algorithm   & Number of Multiplications\\ \hline
		OMP\cite{dncnn}        & $T\tau_P^{\rm DL}KNN_{\rm AP}N_{\rm UE}+L(2T\tau_P^{\rm DL}L^2 + L^3+2T\tau_P^{\rm DL}LK)$                                                   \\ \hline
		AMP\cite{kmltsp}     & $L(20(T\tau_P^{\rm DL}+1)KNN_{\rm AP}N_{\rm UE}+3T\tau_P^{\rm DL}K)$                                                   \\ \hline
		CSINet\cite{csinet}      & $T\tau_P^{\rm DL}KN_{\rm Ker}^2\left[ 64+M_{\rm enc}\left( C_{\rm in}C_{\rm h1}+ C_{\rm in}C_{\rm h1}+C_{\rm h1}C_{\rm h2}\right) \right]+T\tau_P^{\rm DL}KNN_{\rm AP}N_{\rm UE}$                                                   \\ \hline
		Transformer\cite{attn} & $M_{\rm enc}\left[ (2N_{\rm emb}+2)T^2(\tau_P^{\rm DL})^2K^2+4N_{\rm emb}^2 T\tau_P^{\rm DL} K+2N_{\rm emb}N_{\rm hid}T\tau_P^{\rm DL}K  \right]+T\tau_P^{\rm DL}KNN_{\rm AP}N_{\rm UE}N_{\rm emb}$                                                   \\ \hline
		MLP-Mixer\cite{mlpmixer}   & $2T\tau_P^{\rm DL}KN_{\rm emb}[1+M_{\rm enc}(N_{\rm tok}+N_{\rm hid})/(N_{\rm H}N_{\rm W})]+T\tau_P^{\rm DL}KNN_{\rm AP}N_{\rm UE}N_{\rm emb}$                                                   \\ \hline\hline
	\end{tabular}
\end{table*}

	To evaluate the feasibility in practical applications, it is important to analyze the complexity of the proposed methods. Since multiplications are the dominant operation in the proposed methods, we will analyze the number of multiplications of conventional methods\cite{FDD,kmltsp}, learning-based models\cite{csinet,attn,mlpmixer}, and proposed method to evaluate the complexity.
	
	The number of multiplication operations to estimate the downlink channel ${\bf h}_u^{\rm G,DL}$ in Eq.~\eqref{eq:pilottrain4} is given in Table~\ref{tab:complexity}, where $L$ denotes the number of iterations in OMP and AMP algorithms. For CSINet model, $N_{\rm Ker} = 3$ denotes the size of convolutional kernels, $C_{\rm in}=16$, $C_{\rm h1}=64$, and $C_{\rm h2}=32$ respectively denote the numbers of convolutional kernels as shown in Fig.~\ref{fbmodel}(b)(2), and $M_{\rm enc}$ denotes the number of layers, which is identical in all learning-based models. As we can see in the table, all learning-based methods have $L$-independent numbers of multiplications, which shows superiority over conventional methods. Besides, as the number of channels in CNN-based model increases, the number of multiplications increases polynomially. Thanks to the patching embedding operation and non-attention design in Mixer model, the complexity shows the lowest among most adopted structures.

\section{Numerical Results}

In this section, we compare our proposed CSI acquisition and beamforming methods with some state-of-the-art schemes. We firstly evaluate from the performance of the proposed downlink CSI acquisition, and further investigate the performance of the proposed beamforming methods, where both fully-digital array and hybrid analog-digital array are considered.

\subsection{Simulation Setup}
Without loss of generality, we consider a downlink CF-mMIMO scenario, where $N=16$ APs, each equipped with $N_{\rm AP}=4$ antennas, cooperatively serve $U=16$ UEs with $N_{\rm UE} = 2$ antennas (unless otherwise stated). UEs are randomly distributed in a $1000\:{\rm m}\times 1000\:{\rm m}$ area. The large-scale fading factors of signals transmitted from each AP follows 3GPP technical report\cite{channel3GPP} (outdoor UMi scenario) as ${\rm PL}_{\rm LoS} = 32.4 + 40\log_{10}\left( d \right)+20\log_{10}\left( f_c \right)$ and ${\rm PL}_{\rm NLoS} = \max\left({\rm PL}_{\rm LoS},22.4 + 35.3\log_{10}\left( d \right)+21.3\log_{10}\left( f_c \right)\right)$ with LoS probability
\begin{equation}
	{\rm Pr}_{\rm Los} = \frac{18}{d}+\left( 1-\frac{18}{d} \right)e^{-\frac{d}{36}},
\end{equation}
where $d$ is the distance from AP to the UE (in meters), and $f_c$ is the central carrier frequency (in GHz). Wideband signals are transmitted after OFDM modulation, where we have considered $K=192$ subcarriers with $60\:{\rm KHz}$ spacing each. Carrier frequency is set as $f_c = 28\:{\rm GHz}$, and the noise level can therefore be calculated as $\sigma_n^2 = -100.89\:{\rm dBm}$ with noise figure ${\rm NF}=3\:{\rm dB}$.

For the learning-based schemes, we generate $5,000$ environment samples, which together with $U=16$ UEs constitute the training set with $80,000$ individual CSI samples. $90\%$ of the samples are used to form the training set, while the remaining part forms the validation set. Training procedure lasts for $100$ epochs with a batch size of $512$. For the initial $1,000$ iteration steps, we employ a warm-up\cite{warmup} strategy to raise the learning rate linearly from ${\rm lr}_{\rm min} = 3\times 10^{-7}$ to ${\rm lr}_{\rm max} = 3\times 10^{-4}$. In the following training steps, we also employ a cosine annealing strategy\cite{warmup} to smoothly reduce the learning rate to $0$.
\subsection{Channel Acquisition}
\begin{figure*}[!t]
	\centering
	\vspace{-3mm}
	\hspace{-5mm}
	\begin{minipage}[t]{0.33\linewidth}
		\centering
		\includegraphics[scale=0.3]{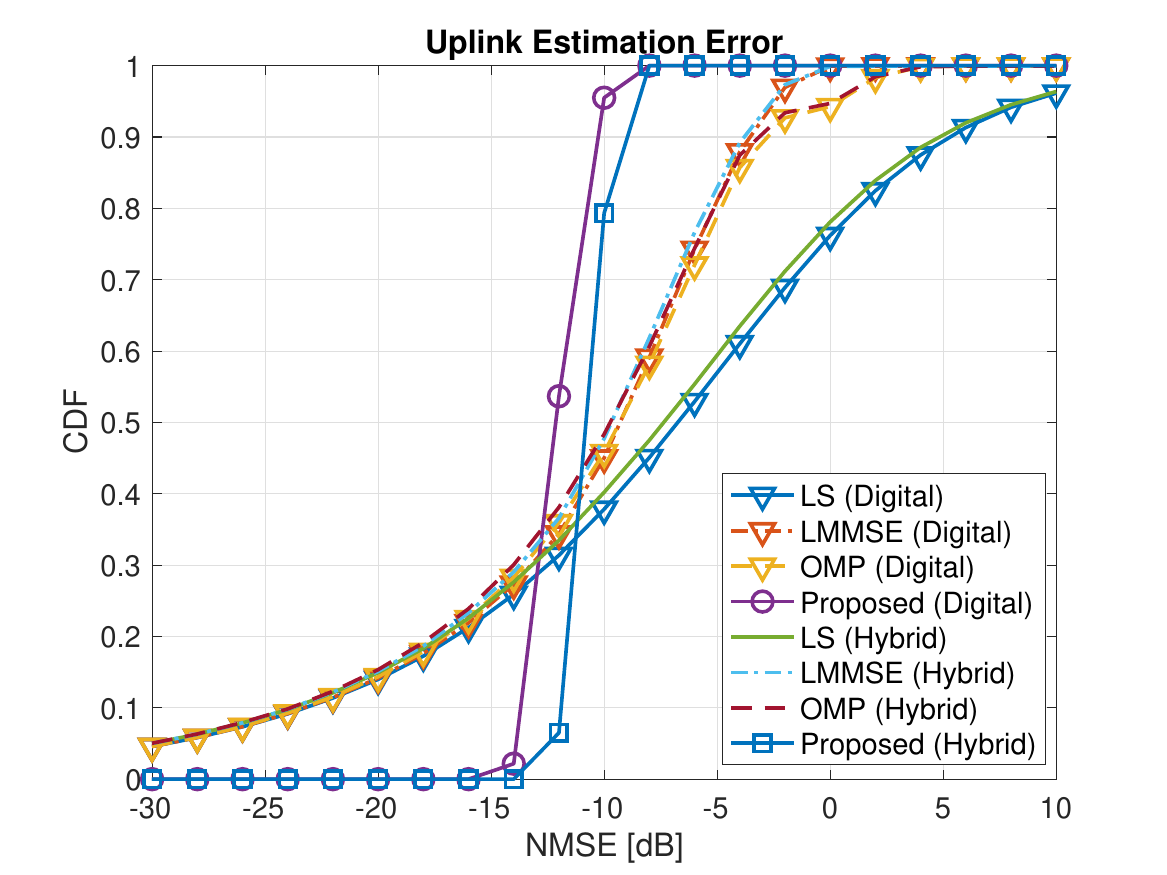}
		\caption{Uplink channel estimation NMSE performance comparison for both fully-digital and hybrid analog-digial array structure.}\label{ulest}
	\end{minipage}
	\hfill
	\begin{minipage}[t]{0.33\linewidth}
		\centering
		\includegraphics[scale=0.3]{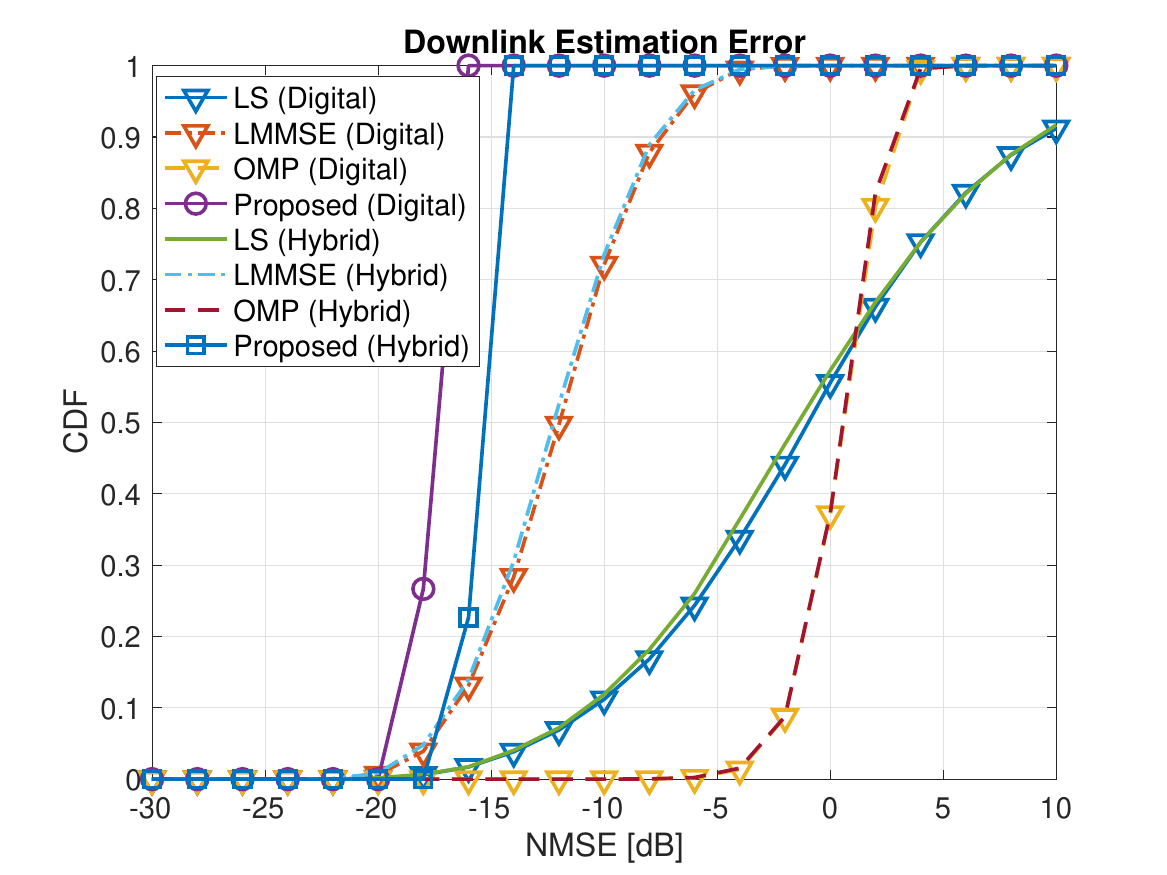}
		\caption{Downlink channel estimation NMSE performance comparison for both fully-digital and hybrid analog-digial array structure.}\label{dlest}
	\end{minipage}
	\hfill
	\begin{minipage}[t]{0.33\linewidth}
		\centering
		\includegraphics[scale=0.3]{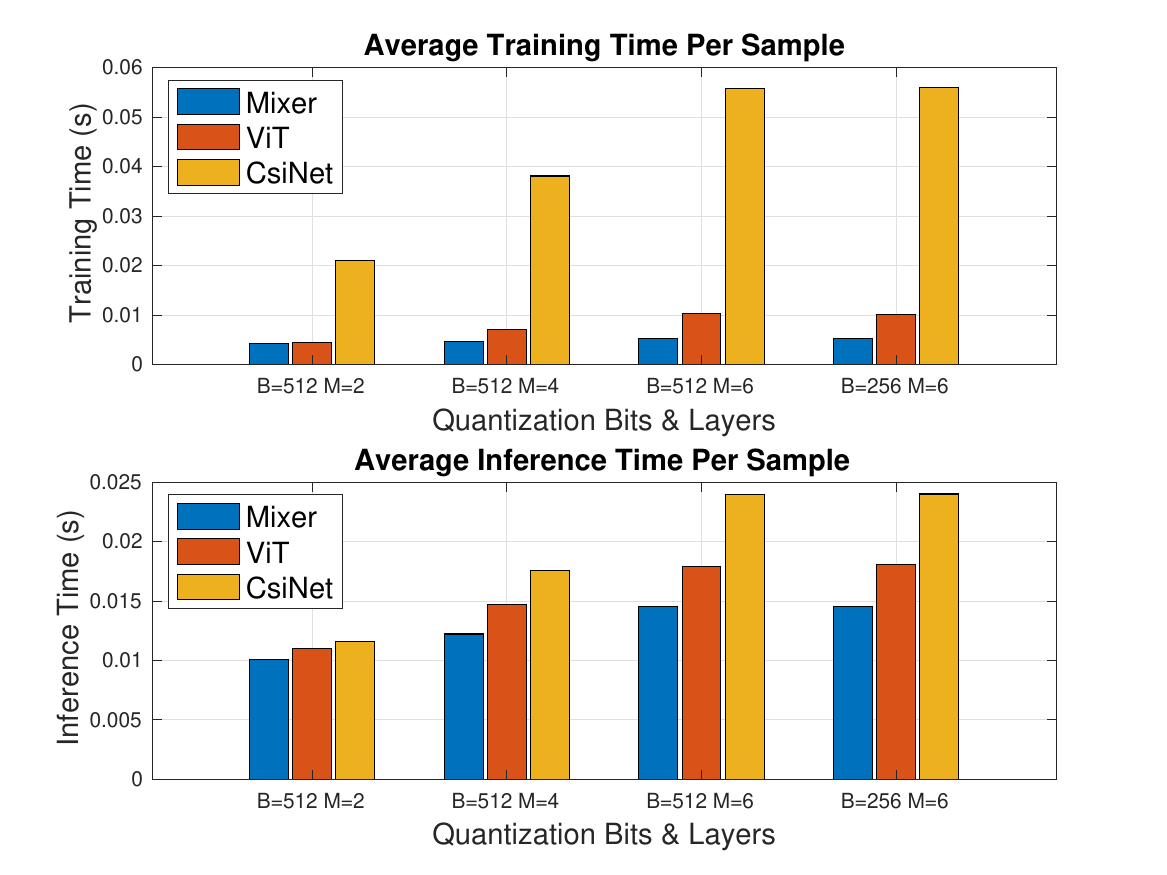}
		\caption{Training and inference time overhead (per sample) with different numbers of layers $M$ and quantization bits $B$.}\label{times}
	\end{minipage}
	\vspace{-3.0mm}
\end{figure*}

We firstly investigate the normalized mean square error (NMSE) performance
\begin{equation}
	{\rm NMSE}=\mathbb{E}\left[ \frac{\Vert \hat{\bf H}-{\bf H} \Vert_F^2}{\Vert {\bf H} \Vert_F^2} \right]
\end{equation}
of our proposed channel estimation schemes. 

For the uplink problem $\rm (P2)$, we set $P_{\rm UE} = 23$ dBm, $\tau_p^{\rm UL} = U N_{\rm UE}$ for both CS-based method\cite{optimality} and linear methods (e.g., LS and LMMSE), while the proposed uplink channel estimation employs $\tau_{\rm AP} = 4$ time slots, i.e., the dimension of measurement of APs. For the proposed uplink estimation method that jointly design the pilot signal and estimator $f^{\rm UL}(\cdot)$, we still adopt the measurements with $\tau_{\rm AP} = 4$ slots at APs, but only transmit pilot signals with half length as $\tau_p^{\rm UL} = \lceil UN_{\rm UE}/2 \rceil$. $M=6$ layers of Mixer modules are used for estimation unless otherwise stated, where the embedding size is $N_{\rm emb} = 192$, and patches are $W\times H = 4\times 6$ in size. $N_{\rm tok} = 256$ tokens are employed, and $N_{\rm hid}=1152$ is set for hidden layers. Cumulative distribution function (CDF) of the NMSE performance for the proposed uplink estimation is shown in Fig.~\ref{ulest}, which depicts that the proposed method not only achieves similar average NMSE performance under half length of pilot signals $\tau_p^{\rm UL}$, but also shows a much more stable performance compared to other state-of-the-art solutions.

For the downlink problem $\rm (P3)$ we set $P_{\rm AP} = 33$ dBm, $\tau_p^{\rm DL} = N N_{\rm AP}$, $\tau_{\rm UE}=N_{\rm UE}$ for LS and LMMSE methods to ensure a full rank measurement for reference, and $\tau_p^{\rm DL} = N N_{\rm AP}/4$ for both CS-based and proposed methods. As illustrated in Fig.~\ref{dlest}, downlink channel estimation schemes show higher performance than uplink scenario thanks to the longer pilot signals and higher transmit power, and our proposed method again shows the best stability of estimation performance. Note that although linear methods show similar performance with our proposed method, the required number of measurements is far larger than our method.

\begin{table}[t]
	\centering
	\caption{Feedback Reconstruction Accuracy with Different $B$, Encoders, and Loss Functions}
	\label{tab:perf}
	\begin{tabular}{cc|cccc}
		\hline\hline
		\multirow{2}{*}{$B$}    & \multirow{2}{*}{Methods} & \multicolumn{2}{c}{FC Encoder} & \multicolumn{2}{c}{Patching Encoder} \\ \cline{3-6} 
		&                          & NMSE             & CSF         & NMSE                & CSF            \\ \hline\hline
		\multirow{7}{*}{64}  & MixerE                   & $-4.650$         & $0.73/0.85$                  & $-4.934$          & $0.75/0.88$                  \\ \cline{2-6} 
		& MixerS                   & $-8.453$         & $0.86/0.92$                  & $-9.168$          & $0.88/0.94$                  \\ \cline{2-6} 
		& MixerUL                  & $\bf -11.311$        & $\bf 0.94/0.96$                  & $\bf -12.346$         & $\bf 0.95/0.98$                  \\ \cline{2-6} 
		& ViT                      & $-7.636$         & $0.83/0.91$                  & $-8.896$          & $0.87/0.93$                  \\ \cline{2-6} 
		& CSINet                   & $-2.655$         & $0.46/0.88$                  & $-2.863$          & $0.49/0.92$                  \\ \cline{2-6} 
		& OMP                      &   $-4.29$               & $0.72$                  &                   &                         \\ \cline{2-6} 
		& AMP                      & $-3.750$                 & $0.67$                  &                   &                         \\ \hline
		\multirow{7}{*}{128} & MixerE                   & $-5.318$        & $0.77/0.87$                  & $-5.745$         & $0.78/0.89$                  \\ \cline{2-6} 
		& MixerS                   & $-10.275$       & $0.91/0.95$                  & $-11.778$        & $0.94/0.97$                  \\ \cline{2-6} 
		& MixerUL                  & $\bf -13.602$       & $\bf 0.96/0.98$                  & $\bf -13.802$        & $\bf 0.96/0.98$                  \\ \cline{2-6} 
		& ViT                      & $-8.892$        & $0.87/0.93$                  & $-9.808$         & $0.90/0.95$                  \\ \cline{2-6} 
		& CSINet                   & $-2.902$        & $0.48/0.90$                  & $-3.001$         & $0.50/0.92$                  \\ \cline{2-6} 
		& OMP                      &   $-7.267$               & $0.82$                  &                   &                       \\ \cline{2-6} 
		& AMP                      &  $-4.725$                & $0.76$                  &                   &                       \\ \hline
		\multirow{7}{*}{256} & MixerE                   & $-6.004$        & $0.78/0.90$                  & $-6.130$         & $0.79/0.90$                  \\ \cline{2-6} 
		& MixerS                   & $-12.927$       & $0.95/0.97$                  & $-13.587$        & $0.96/0.98$                  \\ \cline{2-6} 
		& MixerUL                  & $\bf -14.540$       & $\bf 0.97/0.99$                  & $\bf -14.666$        & $\bf 0.97/0.99$                  \\ \cline{2-6} 
		& ViT                      & $-9.475$        & $0.89/0.95$                  & $-12.945$        & $0.95/0.97$                  \\ \cline{2-6} 
		& CSINet                   & $-3.164$        & $0.53/0.91$                  & $-3.355$         & $0.54/0.92$                  \\ \cline{2-6} 
		& OMP                      & $-7.910$                 & $0.84$                  &                   &                       \\ \cline{2-6} 
		& AMP                      &   $-7.224$               & $0.82$                  &                   &                       \\ \hline
		\multirow{7}{*}{512} & MixerE                   & $-6.284$        & $0.80/0.91$                  & $-6.456$         & $0.81/0.91$                  \\ \cline{2-6} 
		& MixerS                   & $-14.527$       & $0.97/0.98$                  & $-15.825$        & $0.98/0.99$                  \\ \cline{2-6} 
		& MixerUL                  & $\bf -16.071$       & $\bf 0.98/0.99$                  & $\bf -16.250$        & $\bf 0.98/0.99$                  \\ \cline{2-6} 
		& ViT                      & $-14.173$       & $0.96/0.98$                  & $-15.307$        & $0.97/0.98$                  \\ \cline{2-6} 
		& CSINet                   & $-3.540$        & $0.54/0.91$                  & $-3.958$         & $0.55/0.92$                  \\ \cline{2-6} 
		& OMP                      & $-8.172$        & $0.84$                  &                   &                       \\ \cline{2-6} 
		& AMP                      & $-9.079$        & $0.89$                  &                   &                       \\ \hline\hline
	\end{tabular}
\end{table}

As for the problem $\rm (P4)$, we further investigate the performance of the proposed channel SEV module $f_{\rm emb}(\cdot)$ and reconstruction module $f_{\rm rec}(\cdot)$ in reconstructing downlink CSI at APs. Table~\ref{tab:perf} shows the performance of the proposed downlink CSI acquisition from quantized semantic embeddings as shown in Fig.~\ref{fbmodel}(a). 
We use \textit{MixerE} and \textit{MixerS} to distinguish the schemes that estimate CSI before feeding back and directly encoding the pilots into SEVs respectively, and use \textit{MixerUL} to denote the reconstruction scheme using the uplink estimated CSI. Two results are shown in the CSF column in the table, where the left hand side is the CSF evaluated from the model trained by NMSE loss, while the right hand side is trained directly by CSF loss in goal-oriented manner. Here we have considered two kinds of encoders\footnote{For validation purpose, we will try two basic kinds of encoders that are easy to be implemented with low computational complexity, rather than trying to find the ``optimal'' encoder modules.}, namely FC encoder and patching encoder (shown in Fig.~\ref{mixer}), and three kinds of decoders\cite{mlpmixer,attn,csinet}, as we have shown in Fig.~\ref{mixer} and \ref{fbmodel}(b). Mixer, transformer encoder, and CNN-based models have the same number of module layers, i.e., $M=4$, while the other parts of the model as well as the training setup remain unchanged. As for the conventional CS-based methods, we can feedback a subset ${\mathcal{S}^{\rm fb}}$ of the thorough sparse support set ${\mathcal{S}}$ to meet the requirements of $B$, i.e., 
\begin{equation}
	{\mathcal{S}^{\rm fb}}=\underset{\mathcal{S}^\prime \subset \mathcal{S},\vert\mathcal{S}^\prime\vert = N_{\rm fb} }{{\rm argmax}} \sum_{k=1}^K \sum_{s_i\in S}\vert {\bf h}_{s_i}[k]\vert,
\end{equation}
where $N_{\rm fb}=\lfloor \frac{B}{B_{I}+B_{G}}\rfloor$ denotes the number of feedback indices, and $B_I = \log_2 (NN_{\rm AP})$ and $B_G = 32$ denote the quantization bits of indices and corresponding gain, respectively. Performance of feedback reconstruction accuracy is shown in Table.~\ref{tab:perf}, where we can see a significant performance gain in our proposed Mixer-based method, compared with conventional CS-based methods and CNN-based network structure. For feedback scenarios with different numbers of bits, the Mixer-based method with the aid of uplink estimated CSI achieves the best performance, while the performance gain gradually drops as $B$ increases. When we have $B=512\ {\rm bits}$ or even higher feedback overhead, downlink CSI can be well reconstructed barely through the channel semantic embeddings, therefore the aid of uplink CSI is no longer required. 
It is worth mentioning that when the model is trained by CSF loss function, we can see an obvious performance gain especially in the cases that the NMSE loss is high. For example, the reconstruction performance of CNN-based CSINet schemes present a significant gain, which indicates the superiority of the goal-oriented CSF loss function.
\begin{figure*}[t]
	\centering
	\vspace{-3mm}
	\hspace{-5mm}
	\begin{minipage}[t]{0.33\linewidth}
		\centering
		\includegraphics[scale=0.3]{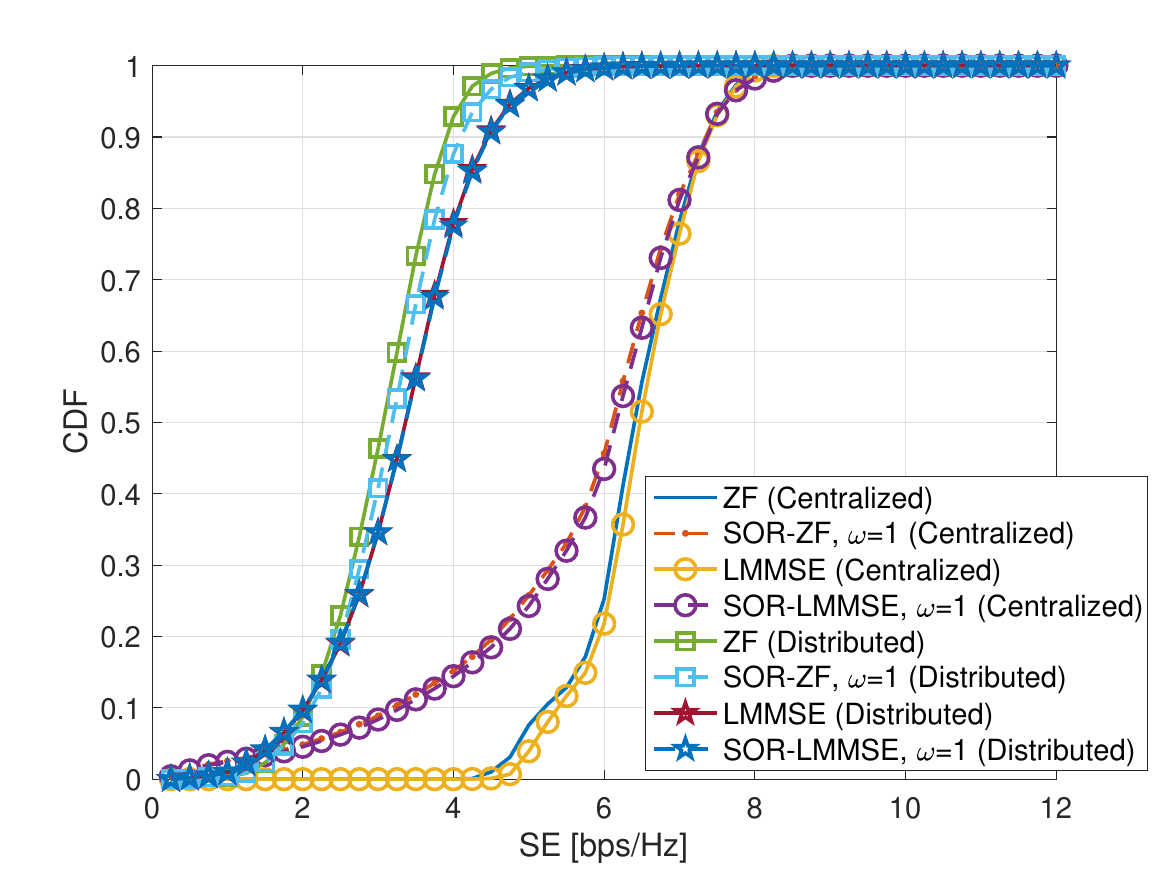}
		\caption{Average SE performance per UE with perfect downlink CSI. SOR-based methods iterate for $L_{\rm max}=10$ times with $\omega=1$.}\label{perfectCSI}
	\end{minipage}
	\hfill
	\begin{minipage}[t]{0.33\linewidth}
		\centering
		\includegraphics[scale=0.3]{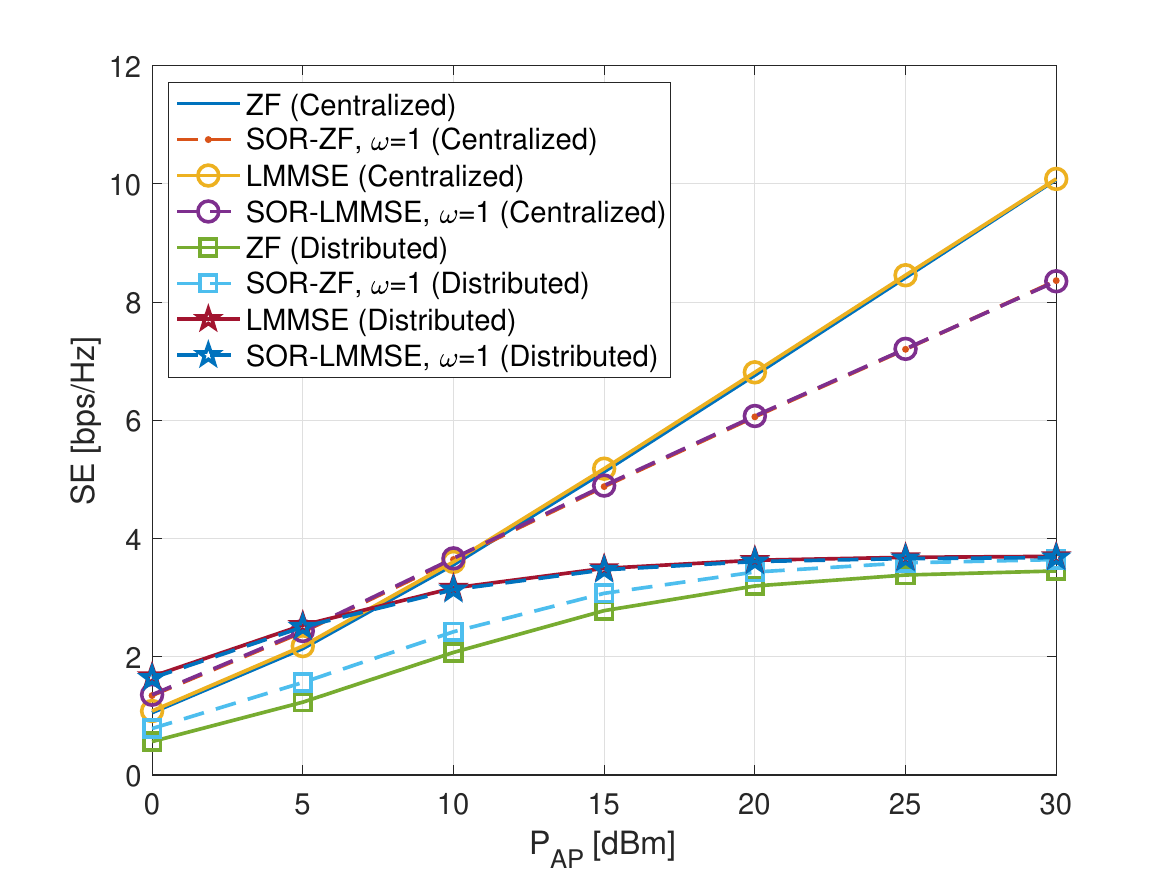}
		\caption{Average SE performance per UE with perfect downlink CSI and varying transmit power $P_{\rm AP}$.}\label{perfectCSI2}
	\end{minipage}
	\hfill
	\begin{minipage}[t]{0.33\linewidth}
		\centering
		\includegraphics[scale=0.3]{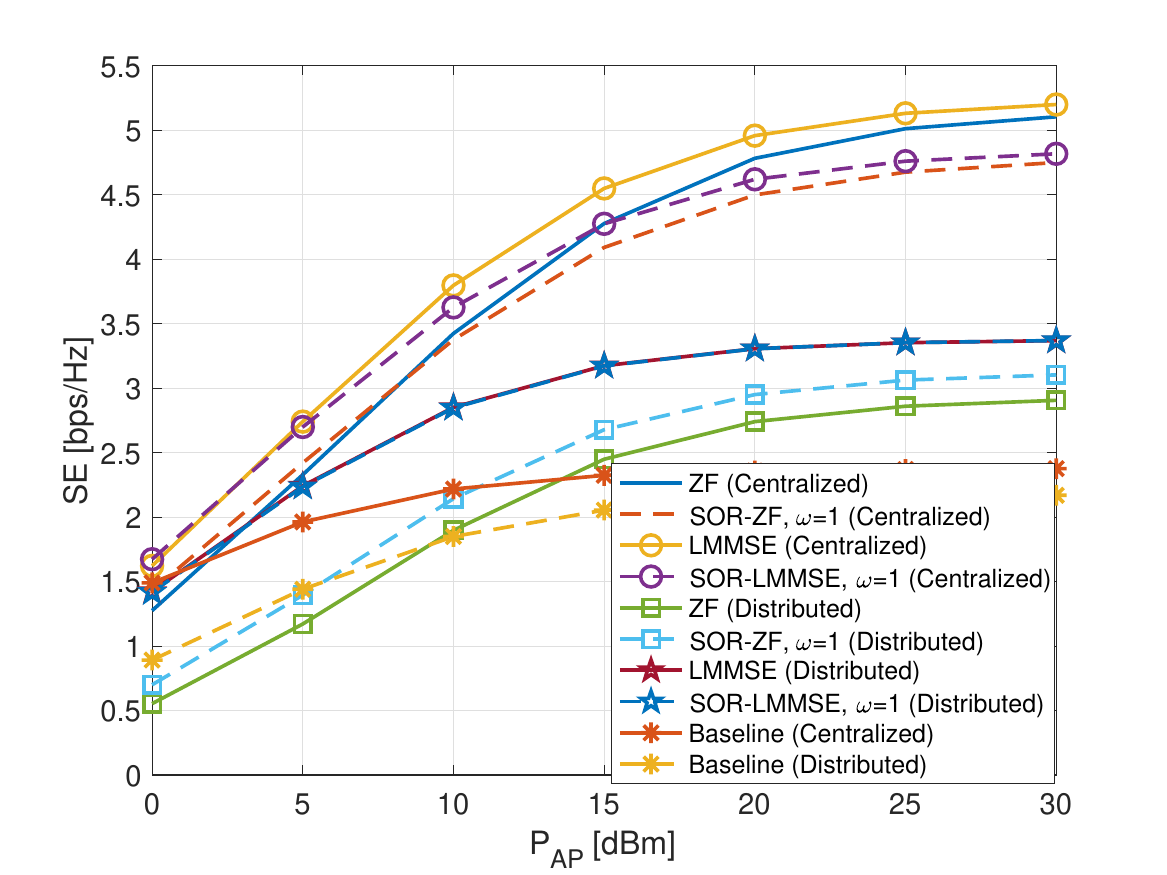}
		\caption{Average SE performance of the proposed digital beamforming per UE with imperfect downlink CSI and varying transmit power $P_{\rm AP}$.}\label{imperfectCSI}
	\end{minipage}
	\vspace{-3.0mm}
\end{figure*}

\begin{figure}[t]
	\centering
	\includegraphics[width=0.45\textwidth]{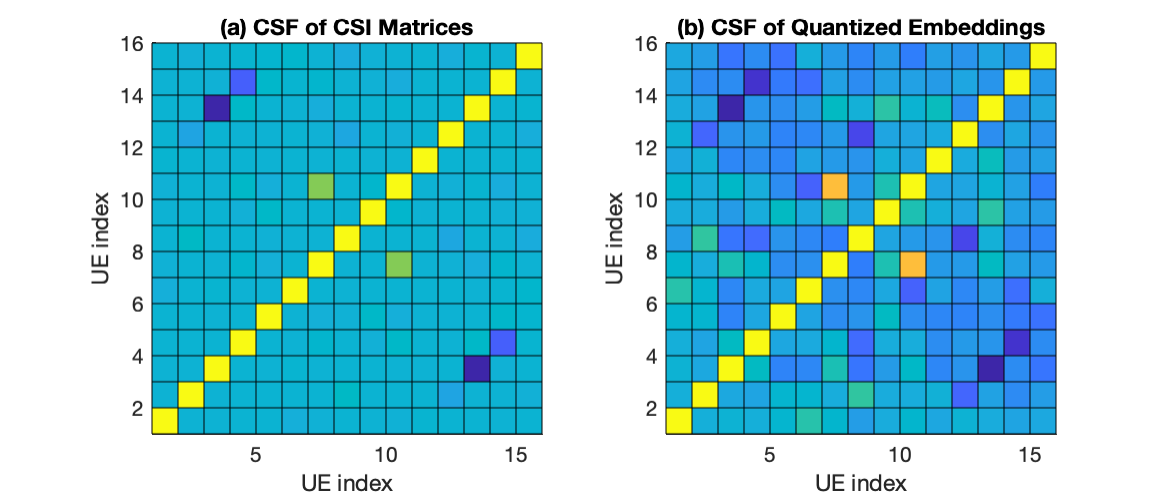}
	\caption{Cosine similarity of (a) downlink channel matrices and (b) quantized low-dimensional embeddings among UEs in one random implementation. }\label{embeddings}
	\vspace{-3.0mm}
\end{figure}

We further investigate the time overhead in both training and inference stages, which is of great importance for practical implementations. As shown in Fig.~\ref{times}, the average time per sample for both training and inference scale up obviously with the number of layers $M$, especially for training cases, while the number of quantization bits $B$ shows insignificant effect. The proposed Mixer-based model shows the best efficiency in both training and inference stages, thanks to its FC layer-based structure and attention-free operation. After the above-mentioned model training is completed, the quantized output of the UE-side encoder constitutes a set of low-dimensional channel embeddings, which can effectively describe the feature of downlink CSI and better distinguish channel matrices from different users, as shown in Fig.~\ref{embeddings}.
\subsection{Deep Unfolding Beamforming Design}

In this section, we will compare the performance of our proposed linear beamforming method with traditional schemes using the reconstructed CSI from feedback semantics. SE performance using perfect CSI is firstly evaluated, while the performance under imperfect CSI with integrated learnable parameters is then provided.

For fully-digital array APs with \textit{perfect CSI} and non-trainable parameter $\omega = 1$, the beamforming performance is given in Fig.~\ref{perfectCSI}, where we have a fixed transmit power $P_{\rm AP}=20\:{\rm dBm}$ for all APs, and the same noise level $\sigma_n^2 = -100.89\:{\rm dBm}$ for all UEs. 
The iterative beamforming schemes are also evaluated with $L_{\rm max}=10$ iterations, unless otherwise stated. The centralized beamforming schemes shows superior performance than distributed schemes with several bps/Hz SE gain, but requiring higher front-hauling overhead. As for the proposed iterative beamforming method, the overall SE performance is very close to the matrix inversion method, which indicates $10$ times of iteration is sufficient for beamforming design. However, the convergence performance of iterative methods is unsatisfactory in centralized processing paradigm. Since the distributed processing paradigm can not effectively mitigate interference from cooperative signal processing, there shows a performance gap between centralized and distributed schemes. The performance with varying transmit power is shown in Fig.~\ref{perfectCSI2}, where the performance of centralized schemes improves as $P_{\rm AP}$ increases, while the performance of distributed schemes gradually approaches a fixed value as $P_{\rm AP}$ increases. The reason for this phenomenon is that when $P_{\rm AP}$ is low, noise level is the dominant factor affecting the performance of distributed scheme. Moreover, when $P_{\rm AP}$ increases, the interference that cannot be eliminated locally becomes the dominant factor affecting the performance.

We further investigate the SE performance of fully-digital beamforming with \textit{imperfect CSI} for centralized and distributed processing paradigms, as shown in Fig.~\ref{imperfectCSI}. We adopt the feedback schemes shown in Table \ref{tab:perf} with $B=64\:{\rm bits}$. Specifically, we perform \textit{MixerS} feedback for centralized processing paradigm, while perform \textit{MixerUL} feedback scheme for distributed processing paradigm to reconstruct local CSI for each AP. We also set two baselines in the performance comparisons, where the feedback model is trained using the NMSE loss rather than CSF loss function.
As shown in Fig.~\ref{imperfectCSI}, imperfect CSI would bring performance degradation for both processing paradigms, and the performance loss for centralized processing paradigm is more severe for two main reasons: (a) the CSI reconstruction accuracy for \textit{MixerS} with the same feedback length under the centralized processing paradigm is much lower than \textit{MixerUL}, and (b) interference cannot be completely eliminated due to imperfect CSI. It is also noted that the performance loss of some iterative-based schemes is relatively lower, which reflects their robustness to imperfect CSI.

\begin{figure*}[t]
	\centering
	\vspace{-3mm}
	\hspace{-5mm}
	\begin{minipage}[t]{0.33\linewidth}
		\centering
		\includegraphics[scale=0.3]{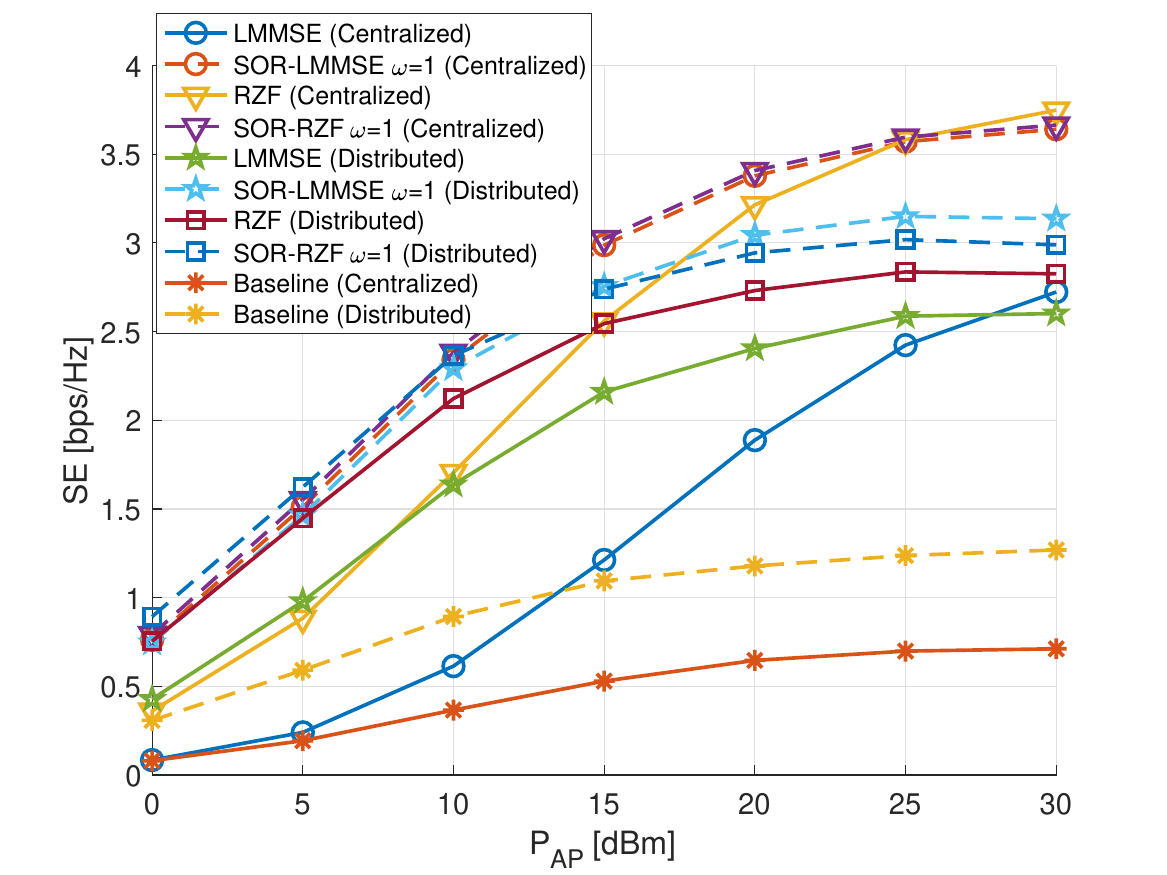}
		\caption{Average SE performance of the proposed hybrid beamforming per UE with imperfect downlink CSI and varying transmit power $P_{\rm AP}$.}\label{hybrid1}
	\end{minipage}
	\hfill
	\begin{minipage}[t]{0.33\linewidth}
		\centering
		\includegraphics[scale=0.3]{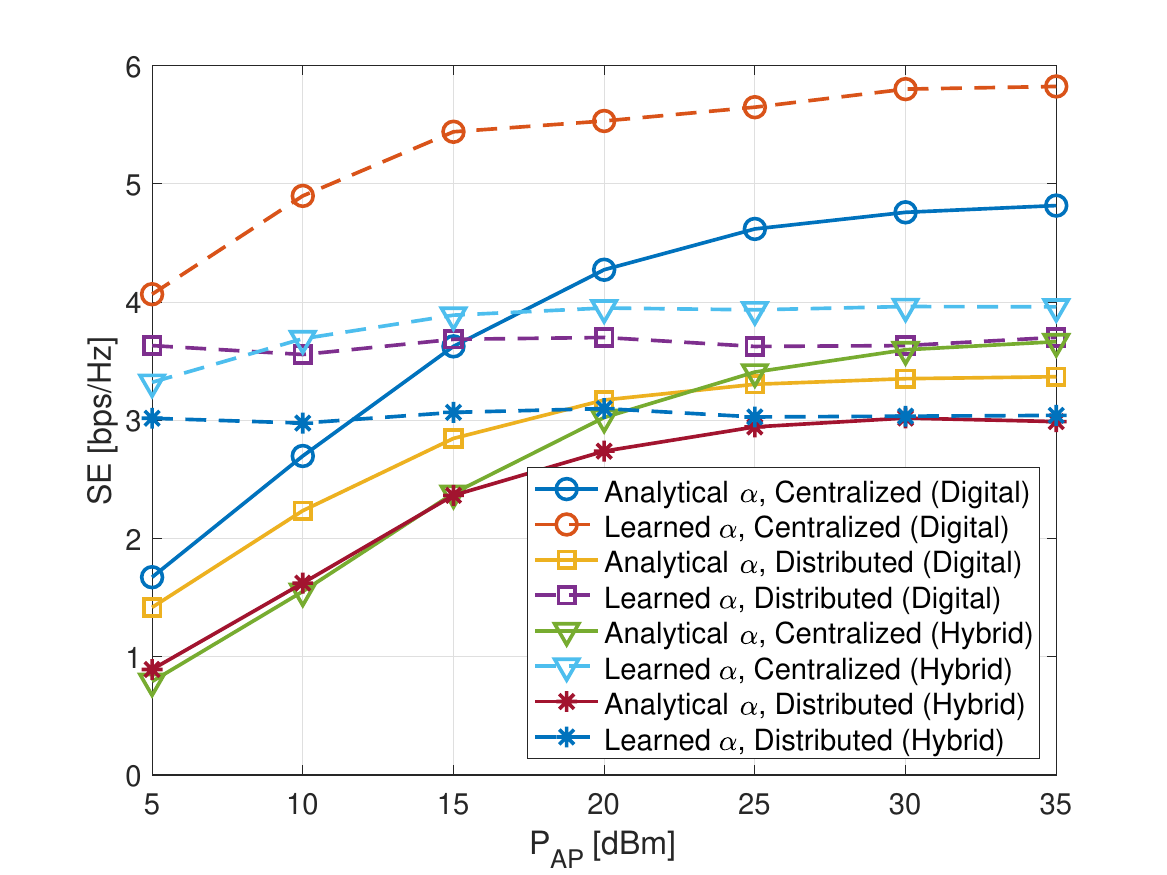}
		\caption{Average SE performance of the proposed deep-unfolding beamforming per UE with imperfect downlink CSI and varying transmit power $P_{\rm AP}$.}\label{learnedrzf}
	\end{minipage}
	\hfill
	\begin{minipage}[t]{0.33\linewidth}
		\centering
		\includegraphics[scale=0.3]{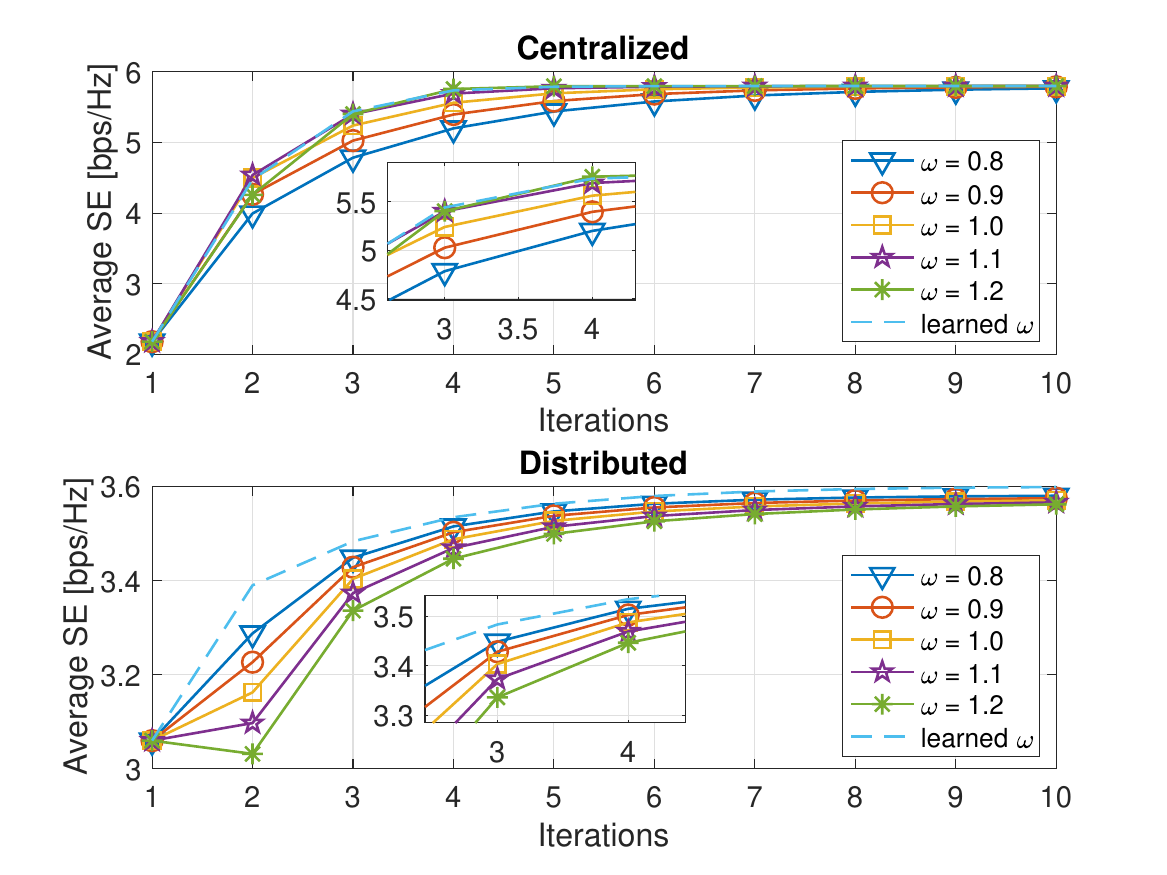}
		\caption{Convergence performance of the proposed deep-unfolding beamforming per UE with imperfect downlink CSI and varying transmission power $P_{\rm AP}$.}\label{converge}
	\end{minipage}
	\vspace{-3.0mm}
\end{figure*}

As for APs with hybrid analog-digital array, the LMMSE and RZF methods are compared in Fig.~\ref{hybrid1}, where the optimal $\boldsymbol{\alpha}$ in RZF-based method is calculated using Eq.~\eqref{eq:optimalrfac}. For distributed processing paradigm, up to $N_{\rm S}=4$ UEs are randomly scheduled to each AP. As the power $P_{\rm AP}$ increases, the increasing feedback accuracy enables the centralized processing paradigm to mitigate interference, which leads to the increasing performance. The close SE performance gap between centralized and distributed schemes also indicates the feasibility of implementing distributed beamforming strategies in practice, whereas the complexity of performing linear beamforming can be further reduced. However, the ``optimal'' $\alpha$ for RZF-based methods shows insignificant effect on improving SE performance here.

\subsection{Learning-based Parameter Selection}

As shown in Fig.~\ref{hybrid1}, the optimal regularization parameter $\alpha$ according to Eq.~\eqref{eq:optimalrfac} is not suitable in the proposed scenario. We hence investigate the performance of the proposed deep-unfolding methods that adaptively learn the optimal parameters via training. Performance comparison is shown in Fig.~\ref{learnedrzf}, where the ``optimal'' regularization factors can be obtained according to the analytical solution Eq.~\eqref{eq:optimalrfac}, and the ``learned'' regularization factors are optimized by the proposed Algorithm 1 and 2. The proposed learning based regularization factor optimization scheme shows significant performance gain especially under low $P_{\rm AP}$ situations, and preserves the performance superiority as $P_{\rm AP}$ grows.

Except regularization factors $\bf \alpha$ and $\bf C$, we also show the effect for optimizing the convergence factor $\omega$. Fig.~\ref{converge} takes digital array architecture as an example, where the convergence performance is compared with manually selected factors around $\omega = 1$. The learned factor $\omega$ shows approximately $96\%$ of the SE performance after merely $3$ iterations of calculation under both centralized and distributed processing paradigms, and it is worth mentioning that our proposed method achieves higher SE performance after sufficient iterations in the distributed processing paradigm, compared with the empirically selected parameters.

\section{Conclusion}

In this paper, we have proposed a hybrid knowledge-data driven CSI acquisition and multi-user beamforming scheme for CF-mMIMO systems to support outdoor XR applications. Specifically, we firstly investigate the channel estimation problem for both uplink and downlink systems, and design a goal-oriented channel semantic embedding scheme for CSI feedback. The proposed scheme surpasses traditional CS-based and learning-based schemes with lower overhead, and has lower training and inference time than state-of-the-art solutions. Furthermore, a model-driven deep-unfolding beamforming method is proposed to adaptively learn the optimal parameters for the low-complexity linear beamforming schemes, which show better robustness to imperfect CSI, and converges faster after training. We also extend our beamforming framework to APs with hybrid analog-digital array, and the performance is close to convergence after merely $3$ iterations, substantially improving the performance under imperfect CSI.

\end{document}